\documentclass[apj]{emulateapj}
\usepackage{graphics,enumerate,amsmath, morefloats}

\newcommand{\niw}{\mbox{[\ion{N}{1}] $\lambda$5197,5200}}

\newcommand{\niibw}{\mbox{[\ion{N}{2}] $\lambda$6583}}

\newcommand{\siiw}{\mbox{[\ion{S}{2}] $\lambda \lambda$6716,6731}}

\newcommand{\oiiibw}{\mbox{[\ion{O}{3}] $\lambda$5007}}

\newcommand{\oi}{\mbox{[\ion{O}{1}]}}
\newcommand{\oii}{\mbox{[\ion{O}{2}]}}
\newcommand{\nii}{\mbox{[\ion{N}{2}]}}
\newcommand{\hal}{\mbox{H$\alpha$}}
\newcommand{\sii}{\mbox{[\ion{S}{2}]}}
\newcommand{\hb}{\text{H$\beta$}}
\newcommand{\oiii}{\text{[\ion{O}{3}]}}

\shorttitle{Nature of LINER-like Emission}
\shortauthors{Yan et al.}

\begin{document}
\title{The Nature of LINER-like Emission in Red Galaxies}
\author{Renbin Yan$^{1}$, Michael R. Blanton$^{1}$} 

\affil{$^1$ Center for Cosmology and Particle Physics, Department of Physics, New York University, New York, NY, 10003; ry9@nyu.edu}

\begin{abstract}
\noindent 

Passive red galaxies frequently contain warm ionized gas and have
spectra similar to low-ionization nuclear emission-line regions
(LINERs). Here we investigate the nature of the ionizing sources
powering this emission, by comparing nuclear spectroscopy from the
Palomar survey with larger aperture data from the Sloan Digital Sky
Survey. We find the line emission in the majority of passive red
galaxies is spatially extended; the \hal\ surface brightness profile
depends on radius $r$ as $r^{-1.28}$. We detect strong line ratio
gradients with radius in \nii/\hal, \sii/\hal, and \oiii/\sii,
requiring the ionization parameter to increase outwards. Combined with
a realistic gas density profile, this outward increasing ionization
parameter convincingly rules out AGN as the dominant ionizing source,
and strongly favors distributed ionizing sources. Sources that follow
the stellar density profile can additionally reproduce the observed
luminosity-dependence of the line ratio gradient. Post-AGB stars
provide a natural ionization source candidate, though they have an
ionization parameter deficit.  Velocity width differences among
different emission lines disfavor shocks as the dominant ionization
mechanism, and suggest that the interstellar medium in these galaxies
contains multiple components.  We conclude that the line emission in 
most LINER-like
galaxies found in large aperture ($>100$ pc) spectroscopy is not
primarily powered by AGN activity and thus does not trace the AGN
bolometric luminosity.  However, they can be used to trace warm gas in
these red galaxies. %These results await proof by future integral field
%spectroscopy observations.

\rightskip=0pt
\end{abstract}

\keywords{galaxies:active --- galaxies: ISM --- galaxies: elliptical and lenticular --- galaxies: emission lines --- ISM: kinematics and dynamics --- stars: AGB and post-AGB}
%
%Introduction
%Data
%Method
%Result
%  1. Velocity width ratios
%  2. Coadded spectra for weak line detection
%  
%Discussion
%  1. List the potential explanation for the line ratio vs. width and
  	
%  References:

%Filippenko & Halpern 1984 : NGC 7213: A key to the nature of LINERs
%Ho, Filippenko & Sargetn 1996:  M81, LINER, linewidth, photo-ionization paper.
%Eracleous et al. 2010 LINER energy buget paper

%Kwok cits Renzini 1982 conference proceeding for lazy AGB star, the paper is actually published in 1983,IAUS, 103, 267 .  "Red giants as precursors of planetary nebulae" But in this paper, Renzini cites his 1981 conference proceeding for lazy AGB stars, which I can only find the abstract. Renzini, A. 1981a, Physical processes in Red Giants, p 431. 
% http://adsabs.harvard.edu/abs/1981ASSL...88..431R

%summary of late stage evolution: 
%Greggio & Renzin, 1990: http://adsabs.harvard.edu/abs/1990ApJ...364...35G

%Lack of post-AGB stars: 
%Brown,T. et al. 2000 : http://adsabs.harvard.edu/abs/2000ApJ...532..308B
%Brown,T. et al. 2008 : http://adsabs.harvard.edu/abs/2008ApJ...682..319B
%Weston et al. arXiv:1010.5408

%Maybe read blue hook population, but they are not hot enough
%http://arxiv.org/abs/1006.1591

\section{Introduction}

Emission lines are important spectral features that can help us probe
the gaseous component in galaxies. They are not unique to star-forming
galaxies, but also exist in galaxies with only old stellar
populations. Numerous results \citep{Phillips86, Goudfrooij94, Yan06}
have shown that line emission is prevalent in more than 50\% of
passive red galaxies, and they have line ratios similar to the low
ionization nuclear emission line regions (LINERs,
\citealt{Heckman80}). What powers this line emission has been an
unsettled question for decades.

LINERs are identified by their particular pattern of line strength
ratios, with strong low-ionization forbidden lines (e.g. \nii, \sii,
\oii, \oi) relative to recombination lines (e.g. \hal, \hb) and
high-ionization forbidden lines (e.g. \oiii). Unlike galaxies
dominated by star-forming HII regions or Seyferts, whose line ratio
patterns clearly identify their ionizing sources as young massive
stars or active galactic nuclei (AGN), respectively, LINERs can be
produced by a wide array of ionization mechanisms, such as
photo-ionization by an AGN \citep{FerlandN83,HalpernS83,GrovesDS04II},
photoionization by post-AGB stars \citep{Binette94}, fast radiative
shocks \citep{DopitaS95}, photoionization by the hot X-ray-emitting
gas \citep{VoitD90, DonahueV91}, or thermal conduction from the hot
gas \citep{SparksMG89}. Therefore, their exact ionization mechanism
has been hotly debated.

The LINER puzzle is futher complicated by the limited spatial
resolution available in many samples, particularly in
SDSS. Originally, ``LINER'' only referred to a class of galaxy
nuclei. They were first identified in {\it nuclear} spectra of nearby
galaxies \citep{Heckman80}. 
This is the case in most LINER studies of
nearby galaxies \citep[e.g.][]{HoFS97V}. For our discussion, we refer
to these LINERs as ``nuclear LINERs.''  
We should keep in mind that ground-based slit spectra, under typical seeing,
usually cannot resolve better than the central few hundred parsecs for
even nearby galaxies ($\lesssim40$ Mpc). This is the scale referred to by the word 'nuclear'.  With narrow band imaging
and/or long-slit spectroscopy surveys, \cite{Phillips86},
\cite{Kim89}, \cite{Buson93}, \cite{Goudfrooij94}, \cite{Macchetto96},
\cite{Zeilinger96}, and others found that the line emission in
early-type galaxies is often extended to kpc scale and has 
LINER-like line ratios.
We refer to these cases as ``extended LINERs.''  In surveys of much
more distant galaxies, such as most galaxies in SDSS, or surveys at
high redshifts, such as DEEP2 \citep{Davis03}, zCOSMOS \citep{Lilly07}, BOSS \citep{Eisenstein11}, and distant cluster surveys 
(e.g. \citealt{Lubin09}), the spectra obtained usually covers a much larger scale than the nuclei and we are not always able to tell how the emission is
distributed spatially. Still, a large number of galaxies show
LINER-like spectra \citep{Yan06,Lemaux10,Bongiorno10,Yan11}. The term ``LINER'' is often casually adopted in
this case to refer to galaxies with LINER-like spectra. Here, we refer
to these cases as ``LINER-like galaxies.''

Although the name of LINER includes a morphological description --- ``nuclear'',
there has been no quantitative definition of what line emission
distribution would qualify as a nuclear LINER. All LINERs have been defined
only spectroscopically based on their line ratio patten. The distinction
between nuclear LINERs and extended LINERs is very murky. Practically, it
often refers to the scale over which the spectrum is taken --- 
$\lesssim200~{\rm pc}$ for nuclear and $\gtrsim1~{\rm kpc}$ for extended, 
rather than a characteristic scale or morphological description of 
the line emission distribution.
\cite{Masegosa11} tried to classify the different morphologies of 
nuclear line emission distribution based on narrow band HST images. 
In the large majority of LINERs (84\%), they found an unresolved nuclear 
source surrounded by diffuse emission extending to a few hundreds of parsecs. 
However, it is unclear in those data whether the two components have different 
ionizing sources and which component dominates in total luminosity.

For a large fraction of nuclear LINERs, which are defined using 
spectra from the central few hundred parsecs, AGN activity clearly exists in
the center, although some puzzles still remain. Evidence for AGN
activity includes the detection of central hard X-ray point sources,
compact radio cores, broad emission lines in direct or polarized
light, and UV variability (see \citealt{Ho08} and references
therein). On the other hand, there is growing evidence
\citep{Ho01,Eracleous10} that the weak AGN in most nuclear LINERs does
not emit enough photoionizing photons to account for the observed
intensity in optical emission lines. Based on narrowband images or
slit spatial profiles observed from HST, the narrow line region in
LINERs appear to be strongly concentrated in the center, with typical
dimensions smaller than tens of parsecs \citep{Walsh08}. However, in
some objects the profile can extend to a few hundred parsecs
\citep{Shields07}. Thus, even for nuclear LINERs, whether the AGN is
responsible for all of the narrow line emission within the central few
hundred parsecs is unclear.

For the extended LINERs, it is totally unclear how they are related with 
nuclear LINERs and what mechanism powers their line emission. For LINER-like 
galaxies, we know even less. Are they very powerful nuclear LINERs or are they 
dominated by extended LINER emission? 
The nuclear-LINER fraction among early-type galaxies is fairly similar to the 
LINER-like galaxy fraction among red galaxies in SDSS \citep{HoFS97V, Yan06}.
Is this similarity fortuitous, or is there a physical connection? We try to answer these questions in this paper.

Solving these puzzles is important. Many people have used the line
strength in LINER-like galaxies from SDSS as an indicator of AGN power
\citep{KauffmannHT03,Constantin06, Kewley06, Schawinski07,
  KauffmannH09}, while others argued they are not genuine AGNs
\citep{Stasinska08, Sarzi10, CidFernandes11, Capetti11} but more likely
powered by hot evolved stars. Therefore,
settling this puzzle is not only important for AGN demographics, but
also for understanding the ISM in early-type galaxies. If the line
strength is not an AGN indicator, it might instead reflect the amount
of cool gas supply, or the cooling rate of the hot gas. To make the
correct physical connections, we have to first find out what powers
the line emission.  Additionally, if we can pin down the ionization
mechanism, we will be able to measure the gas-phase metallicity in
these red galaxies, which has so far been impossible.

For the extended LINER emission, three types of evidence have been 
presented to argue that they are not powered by AGN. 
(a) Post-AGBs could produce enough photo-ionizing
photons \citep{dSA90, Binette94, Stasinska08}. (b) The line emission regions 
are spatially extended with a surface brightness profile that is shallower 
than $r^{-2}$ \citep{Sarzi10}. (c) The line
luminosity correlates with stellar luminosity \citep{Sarzi10, Capetti11}. 
This evidence led to the reasonable suspicion that a population of hot 
evolved stars might in fact dominate over the AGN photoionization in 
early-type galaxies. However, all of these arguments depends on inherent 
assumptions about observationally unknown factors.

For the first argument, \cite{dSA90}, \cite{Binette94}, and 
\cite{Stasinska08} all assumed that nearly all the photons produced by 
post-AGB stars are absorbed by the gas and that the gas is distributed near 
the ionizing stars to produce the correct ionization parameter. Both the gas 
covering factor and the relative distribution of gas to stars are unknown. 

For the second argument, \cite{Sarzi06, Sarzi10} showed that the line emission 
in nearly all line-emitting early-type galaxies is spatially extended, and 
has a surface brightness profile shallower than $r^{-2}$. Although 
\cite{Sarzi10} is very careful in not drawing conclusions based on this fact 
alone, such extended emission has been quoted by many others 
\citep[e.g.][]{Kaviraj10,Masters10,Schawinski10} as evidence for stellar 
photoionization. However, a central point source can also produce extended
line emission regions. The line emission surface brightness profile not only
depends on the ionizing flux profile, but also depends on how the 
gas filling factor, spatial distribution of gas clouds, and the gas density 
vary with radius. 
Both the filling factor and the cloud spatial distribution are so poorly known
that the surface brightness profile provides no practical constraint on the 
flux profile.

The third argument is based on the observed correlation in surface brightness 
between line emission and stellar continuum. \cite{Sarzi10} showed 
with IFU data that the \hb\ equivalent width (EW) is fairly constant throughout 
each galaxy when excluding the nuclear region; \cite{Capetti11} argued 
a similar point based on integrated line emission from SDSS.
This might seem like the strongest support for stellar photoionization.  
However, as mentioned above,
the line flux depends on many other unconstrained factors besides the ionizing 
flux, such as the gas filing factor.
\cite{Sarzi10} performed the calculation for a simple model of stellar 
photoionization and found that it did not produce the spatially-invariant 
EW of \hb\ they found throughout each galaxy in their sample. 
To make the model consistent with observations, certain fine tuning of 
the gas filling factor, density, and/or mean-free path of the ionizing 
photons is required, which has no independent observational support. 

In fact, as long as both the line flux and stellar continuum are strong 
functions with radius in these galaxies, one would always find tight 
correlation between the two, even in the case of photoionization by an AGN.

Thus, we need a simpler test that can distinguish different 
ionization mechanisms that is free of assumptions about unknown parameters.
A central point source (e.g. AGN) and a system
of spatially distributed ionizing sources will produce different
ionizing flux profiles as a function of radius. Thus, for the same gas
density profile, they would yield different ionization parameter
profiles, leading to different spatial gradients in line ratios. Thus,
looking at the line ratio gradients may provide a clue to
differentiate the two scenarios. 
The only assumption involved here
is the gas density profile, which can be derived from the hot gas density
profile assuming pressure equilibrium or can be measured directly from line ratios.

Ideally, the line ratio gradient is best measured from integral field 
spectroscopy (IFS) data. However, current IFS data (SAURON, \citealt{Bacon01}; 
ATLAS3D, \citealt{Cappellari11}) have 
very limited wavelength coverage and do not probe enough strong emission 
lines to detect the ionization gradient in a large number of galaxies. 
The emission lines covered by SAURON are \oiii\ , \hb, and in a few 
cases \niw. Because \oiii/\hb\ depends on the ionization parameter and to some extent also on the hardness of the ionizing spectra, it alone does not provide an unambiguous constraint on the ionization gradient. In addition, the weakness of \hb\ makes it more difficult to detect small changes.

Our solution is to use the nuclear spectra from the Palomar survey \citep{HoFS95},
and the fiber spectra from SDSS. By identifying the same population of
galaxies at different redshifts, for which the fixed angular aperture
corresponds to different physical radii, we can study statistically
the spatial profile of emission line surface brightness and the line
ratio gradient. The emission line surface brightness profile can also
inform us about the relationship between nuclear LINERs and extended
LINERs.

In addition, we will examine the widths of emission lines.  Different
forbidden lines will have different widths if there is line ratio
variation within a galaxy and the variation is correlated with gas
kinematics. 
On the other hand, shock ionization models also should produce width
differences among multiple lines due to the dependence of line ratios
on shock velocity.

\begin{figure}
\begin{center}
\includegraphics[totalheight=0.35\textheight]{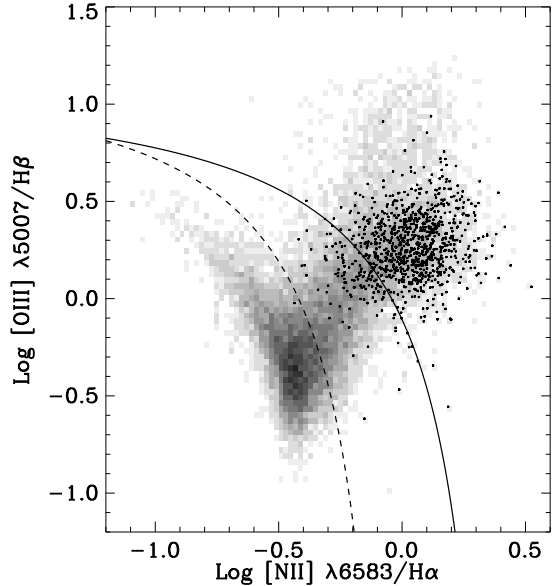}
\caption{Line ratio diagnostic diagram for SDSS galaxies (gray scale) at $0.09<z<0.1$ with all four emission lines detected at more than 3$\sigma$ significance, and the passive red galaxies (black points) among them. The latter is selected according to the criteria described in Section~\ref{sec:sampleselect}.
The curves represent demarcations defined by \cite{KewleyDS01} (solid) and \cite{KauffmannHT03} (dashed). This illustrates that red galaxies have fairly uniform line ratios which puts most of them in the LINER-like galaxy category. 
}
\label{fig:bpt}
\end{center}
\end{figure}

Our investigation includes all line-emitting red galaxies except for
those containing star formation; we do not specifically select for
LINER-like galaxies.  Based on line ratio diagnostics, these
line-emitting red galaxies do have fairly uniform line ratios with the
most typical case belonging to the LINER category, as shown by the
line ratio diagnostic diagram in Figure~\ref{fig:bpt}.  Avoiding the
use of line ratios in sample selection is essential for our study of
the line ratio gradient.

This paper is organized as following. In Section 2, we will describe
the data, measurements, and sample selection. In Section 3, we will
investigate the relationship between nuclear LINERs and extended LINERs
and derive an average emission line surface brightness profile. In Section 4,
we will show the line ratio gradient. In Section 5, we will present
the line width differences. In Section 6, we discuss the viability of
each ionization mechanism in explaining the line ratio gradient and
line width differences. We conclude in Section 7.

Throughout this paper, we use a flat $\Lambda$CDM cosmology with $\Omega_m=0.3$,
and a Hubble constant of $H_0=75 h_{75}{\rm km s^{-1} Mpc^{-1}}$ with $h_{75}=1$. This Hubble constant is chosen to make the comparison easier with the data presented by \cite{HoFS97III}. All the magnitude used are in the AB system.

%Critical density:
%[O III] 4363      : 3  x10^7
%[O I]   6300      : 2  x10^6
%[O III] 5007, 4959: 6.8x10^5
%[N II] 6583,6548:   6.6x10^4
%[O II] 3729:        3.4x10^3   (2D5/2)
%[O II] 3726:        1.5x10^4   (2D3/2)
%[S II] 6731:        2  x10^3   (2D3/2)  3.6x10^3
%[S II] 6716:        2  x10^3   (2D5/2)  1.4x10^3

\section{Data and Measurements}

To investigate the luminosity and line ratio gradient, we compare line
luminosity and line ratio measurements between different physical
apertures for the same population of galaxies. The Palomar survey
\citep{HoFS95} provides us the best sample for nuclear aperture
measurements. The SDSS survey can provide consecutively larger
aperture measurements if we select identical samples at consecutively
higher redshifts.

\subsection{Data}
In the Palomar survey, nuclear spectrum were taken for a sample of
$\sim500$ local galaxies, selected from the Revised Shapley-Ames
Catalog of Bright Galaxies (RSA; \citealt{SandageT81}) and the Second
Reference Catalogue of Bright Galaxies (RC2; \citealt{RC2}) with the
criteria of $B_T<12.5$ (Vega magnitudes) and $\delta >0$. The nuclear
regions ($\sim200$ pc) of these galaxies are isolated using a
$2\arcsec \times4\arcsec$ aperture. The details of data reduction,
stellar continuum subtraction, and line measurements can be found in
\cite{HoFS95}.

We also employ data from the Sloan Digital Sky Survey \citep{York00,
  Stoughton02} Data Release Seven \citep{SDSSDR7}. Using two fiber-fed
spectrographs on a dedicated 2.5-m telescope, SDSS has obtained high
quality spectra for roughly half a million galaxies with $r<17.77$ in
the wavelength range of 3800-9200\AA\ with a resolution of
$R\sim2000$. The SDSS fibers have a fixed aperture of 3$''$ diameter,
which corresponds to different physical scales at different distances.

The SDSS spectroscopic data used here have been reduced through the
Princeton spectroscopic reduction pipeline, which produces the flux-
and wavelength-calibrated
spectra.\footnote{http://spectro.princeton.edu/} The redshift catalog
of galaxies used is from the NYU Value Added Galaxy Catalog (DR7)
\footnote{http://sdss.physics.nyu.edu/vagc/} \citep{BlantonSS05}.
K-corrections for SDSS were derived using \cite{BlantonR07}'s {\it
kcorrect} package (v4\_2). 

\subsection{Emission line measurements}
For emission line measurements in the Palomar sample, we adopt the tabulated values provided by \cite{HoFS97III}. 

For the SDSS sample, we measured the emission lines in the spectra after 
a careful subtraction of the stellar continua. The code used is an updated 
version of the code used by \cite{Yan06}. The major updates are:
\begin{enumerate}
\item We apply an additional flux calibration to all of SDSS spectra
  to fix small scale calibration residuals \citep{Yan11flux}. This is
  critically important for our results. We describe this correction in
  more detail below.
\item The absolute flux is calibrated for each spectrum by matching
  the synthetic $r$-band magnitude with the $r$-band fiber
  magnitude. The spectra are also corrected for Galactic extinction.
\item The stellar continuum is modelled as a non-negative linear
  combination of 7 templates. The templates are seven simple stellar
  population models with solar metallicity, with ages of 0.125, 0.25,
  0.5, 1, 2, 7, and 13 Gyrs, made using the \cite{BC03} stellar
  population models.
\end{enumerate}

As in \cite{Yan06}, the emisson line flux is measured from direct flux-summing in the continuum-subtracted spectra. The line windows and sidebands are unchanged.

We discovered that the flux calibration produced by the standard SDSS
pipeline has percent-level small-scale residuals that can
significantly impact the measurement of emission line flux when the
equivalent width of the line is low (a few Angstroms; see \citealt{Yan11flux} 
Fig.~1 for an example of the impact). For example, for an emission
line EW of 1\AA\ measured in a 20\AA window, if the throughput differs
by 1\% between the central window and the sidebands where the
continuum level is measured, the line flux measurement will be off by
20\%. This will introduce systematic offsets in line flux and line
ratios as a function of redshift, significantly hampering our
investigation. Therefore, we need a much more accurate small scale
flux calibration than what the standard pipeline produces.

\cite{Yan11flux} solved this problem by comparing stacked red-sequence
spectra between small redshift intervals to statistically determine
the relative throughput as a function of wavelength, and achieved an
flux calibration accurracy of 0.1\% on scales of a few hundred
Angstrom. We applied this small scale flux calibration to the
spectra. This calibration is essential for the result presented in
this paper (see Fig. 6\ in \citealt{Yan11flux}).

In this paper, we also make use of the line width measurements for the
SDSS sample. The line widths are measured by fitting Gaussians to each
emission line. Different emission lines are allowed to have different
widths, except that the two \sii\ lines (6716\AA\ and 6731\AA) and the
two \nii\ lines (6548\AA\ and 6584\AA) are both forced to have the
same width. The instrumental resolution of SDSS varies with
wavelength, the position of the fiber on the focal plane, and the
spectrograph. This varying resolution as a function of wavelength is
given for each individual spectrum by the Princeton pipeline. We
subtracted quadratically the instrumental broadening from the measured
line width to derive the intrinsic width of each emission line for
each galaxy. Our quoted uncertainty of the line width measurement is
the formal uncertainty of the Gaussian fit.

\subsection{Photometry}

To identify the same population of galaxies at different redshifts, we
use a photometric selection. We intentionally avoid the use of line
ratios in sample selection to avoid bias on the line ratio gradient
estimates.

For the Palomar survey, we took the catalog provided by
\cite{HoFS97III}. Photometric information is available from the Third
Reference Catalogue of Bright Galaxies (RC3;\citealt{RC3}) but is
incomplete. To increase the sample size with available photometry and
to put them on the same system as the SDSS galaxies, we re-measured
photometry for those galaxies inside the SDSS footprint using the SDSS
images. We employed an improved background subtraction technique
\citep{Blanton11} to treat these nearby large galaxies properly. After
proper background subtraction, mosaicking, and deblending, we measure
the photometry by fitting a two-dimensional Sersic profile to the
deblended galaxy image. In the end, we derive the Galactic extinction
corrected restframe $B$ and $V$ magnitudes for these galaxies from the
measured $g$ and $r$ magnitudes using the {\it kcorrect} software
package (v4\_2, \citealt{BlantonR07}).

For those galaxies outside the SDSS footprint and for certain Messier
objects for which the new method does not yield satisfactory results,
we take the photometric information from the RC3 catalog and convert
them to the AB system, and then correct for Galactic extinction.  We
do not attempt to correct for internal extinction for these galaxies
as such measurements are not available for the SDSS sample.

For the higher redshift SDSS spectroscopic sample we derived the $B$
and $V$ magnitudes from the SDSS magnitudes using the {\it kcorrect}
package mentioned above.

\subsection{Sample definition} \label{sec:sampleselect}

\begin{figure}
\begin{center}
\includegraphics[totalheight=0.35\textheight]{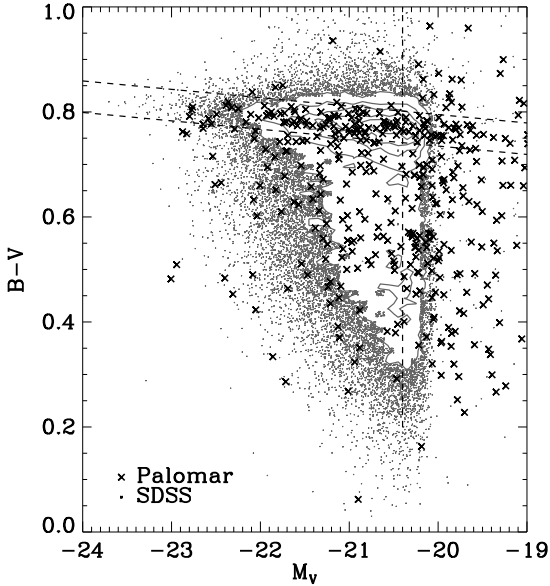}
\caption{Color-magnitude distribution of the Palomar sample (dark crosses) and
galaxies in SDSS with $0.09<z<0.1$ (contour and gray points). The lines
indicate our color and magnitude cuts.}
\label{fig:mb_bv_sloanpalomar}
\end{center}
\end{figure}

Figure~\ref{fig:mb_bv_sloanpalomar} shows the color-magnitude
distribution for the Palomar sample overlaid on a sample of SDSS galaxies 
between $0.09<z<0.1$. The two samples have consistent color-magnitude distributions. We select only the red-sequence galaxies in both samples
using two stringent color cuts defined by 
\begin{align}
  (B-V) &> -0.016(M_V-5\log h_{75})+0.415 \\
  (B-V) &< -0.016(M_V-5\log h_{75})+0.475
\end{align}. 

These cuts are chosen to reduce contamination from dusty star-forming
galaxies.  We limit to galaxies brighter than $-20.4$ in $M_V-5\log
h_{75}$ to match the magnitude limit of the SDSS survey at
$z\sim0.10$.

There are 86 red galaxies in the Palomar survey satisfying these two
cuts. Based on the morphological type given by the RC3 catalog, there
are 30 ellipticals, 30 lenticulars, 15 early spirals(S0/a, Sa, Sab), 9
late spirals (Sb,Sbc,Sc) and 2 irregular galaxies.  With the
classification scheme given in \cite{HoFS97III}, there are 29 LINER
nuclei, 19 transition objects, 11 Seyferts and 3 HII regions. The
remaining 24 objects have no line emission detectable in their nuclei.

In SDSS, we select a comparison sample with $0.01<z<0.1$ using the same
color and absolute magnitude cuts. In most of our analysis, we bin the SDSS
sample into 9 redshift bins with $\Delta z=0.01$.

Despite our selection on color, red galaxies can have sufficient star
formation to contribute to the line emission in our apertures,
especially for the more distant galaxies.  The morphological
distribution of the Palomar red galaxy sample suggests that this is occuring, given the presence of late-type spirals.  We would like to exclude
star-forming spectra in our analysis, since we want to understand the
origin of line emission not associated with star formation. Therefore,
in the Palomar sample, we exclude galaxies with Hubble types later
than S0 and those spectroscopically classified as HII nuclei. This
removes 31\% of the Palomar red sample. The remaining sample includes
19 LINERs, 13 transition objects, 6 Seyferts, and 21 quiescent
galaxies.

\begin{figure}
\begin{center}
\includegraphics[totalheight=0.35\textheight]{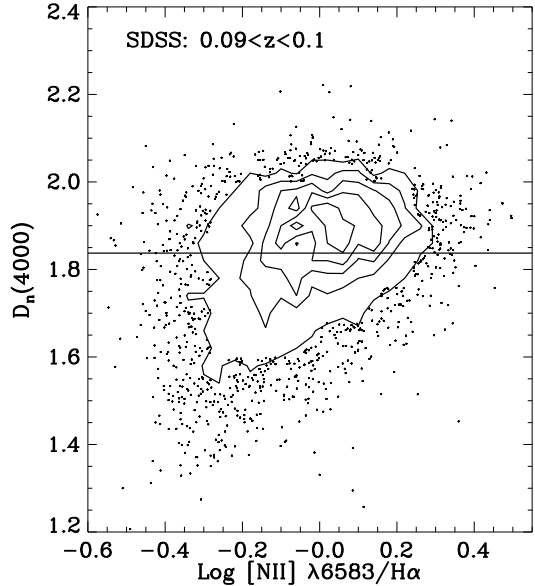}
\caption{$D_n(4000)$ vs. $\log \nii/\hal$ for red sequence galaxies in
  SDSS with $0.09<z<0.1$ and $M_V < -20.4$. We show only the brightest
  50\% of the sample in \hal\ luminosity. We show our chosen threshold
  as the solid horizontal line; it is set at the 30-th percentile in
  the $D_n(4000)$ distribution of the whole sample. Those galaxies
  with low $D_n(4000)$ and low $\nii/\hal$ probably have significant
  contribution by young massive stars in the production of their line
  emission. }
\label{fig:n2ha_d4000_z0.1}
\end{center}
\end{figure}

To achieve a similar selection in the SDSS sample, we exclude galaxies
with any star formation using a stringent cut on $D_n(4000)$
(\citealt{Balogh99}). This quantity is a proxy for the light weighted
mean stellar age, and is thus sensitive to small levels of star
formation. Measured over two 100\AA\ windows separated by 50\AA, it is
less affected by dust reddening than rest frame colors, and more
robustly measured than the H$\delta_A$ equivalent
width. Figure~\ref{fig:n2ha_d4000_z0.1} shows the $D_n(4000)$
vs. $\log \nii/\hal$ for those galaxies in the SDSS sample with
$0.09<z<0.1$. Galaxies that have small $D_n(4000)$ also tend to have
lower \nii/\hal, suggesting that star formation could be contributing
significantly among these. In the SDSS sample, we remove the 30\%
galaxies with the lowest $D_n(4000)$. This selection by $D_n(4000)$
rank is done separately in each redshift bin to take into account any
potential aperture effects and redshift evolution.
We choose a cut on $D_n(4000)$ rather than one based on line ratios to
avoid biasing the comparison of emission line properties.  In the rest
of this paper, we refer to samples with possible star-forming galaxies
removed as the Palomar red sample and the SDSS red sample.

To summarize, from the Palomar survey and SDSS, we identified a volume-limited sample of passive red galaxies without any star formation at $0<z<0.1$.

\section{Spatial distribution of line emission}
In this section, we investigate the spatial distribution of line emission.
The spatial distribution alone does not distinguish between different
ionization mechanisms, but it is essential for the interpretation of
other measurements, such as line ratio gradients. 

We do this in a two-step process. First, we compare the nuclear aperture measurements
from the Palomar survey with the large aperture measurements from SDSS at $z\sim0.1$
to establish the relation between nuclear LINERs in Palomar and the LINER-like galaxies in SDSS. Then, we utilize all apertures available to us from $0<z<0.1$ to measure the average emission line surface brightness profile among passive red galaxies.

\subsection{Nuclear Emission vs. Extended Emission}

In this section, we will compare the emission line luminosity distributions
between two identically-selected, volume-limited samples, but for which 
the line luminosities are measured from different physical apertures. 
The difference in their luminosity distributions demonstrates that 
the emission measured in the larger aperture has to be spatially extended.

\begin{figure*}
\begin{center}
\includegraphics[totalheight=0.35\textheight]{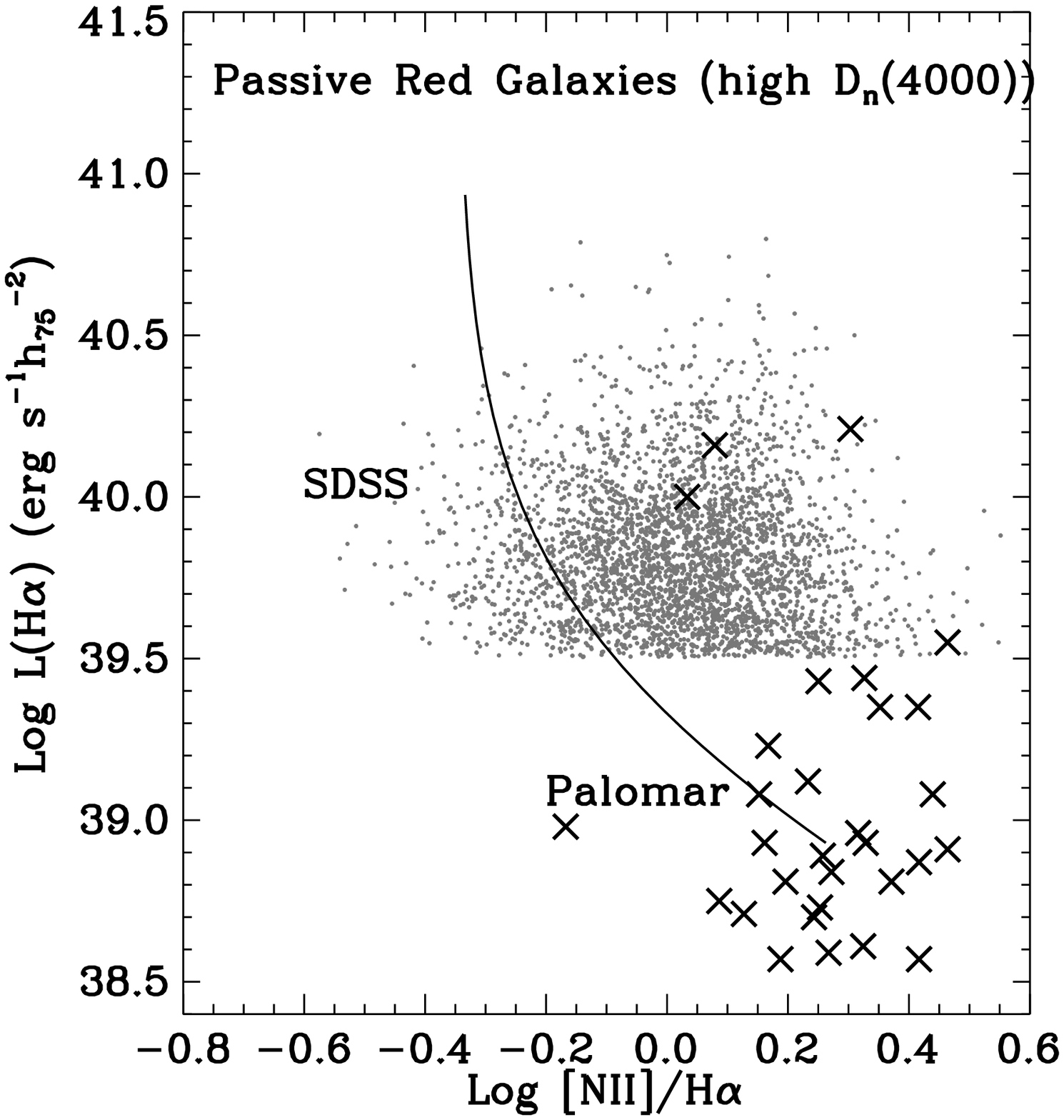}
\includegraphics[totalheight=0.35\textheight]{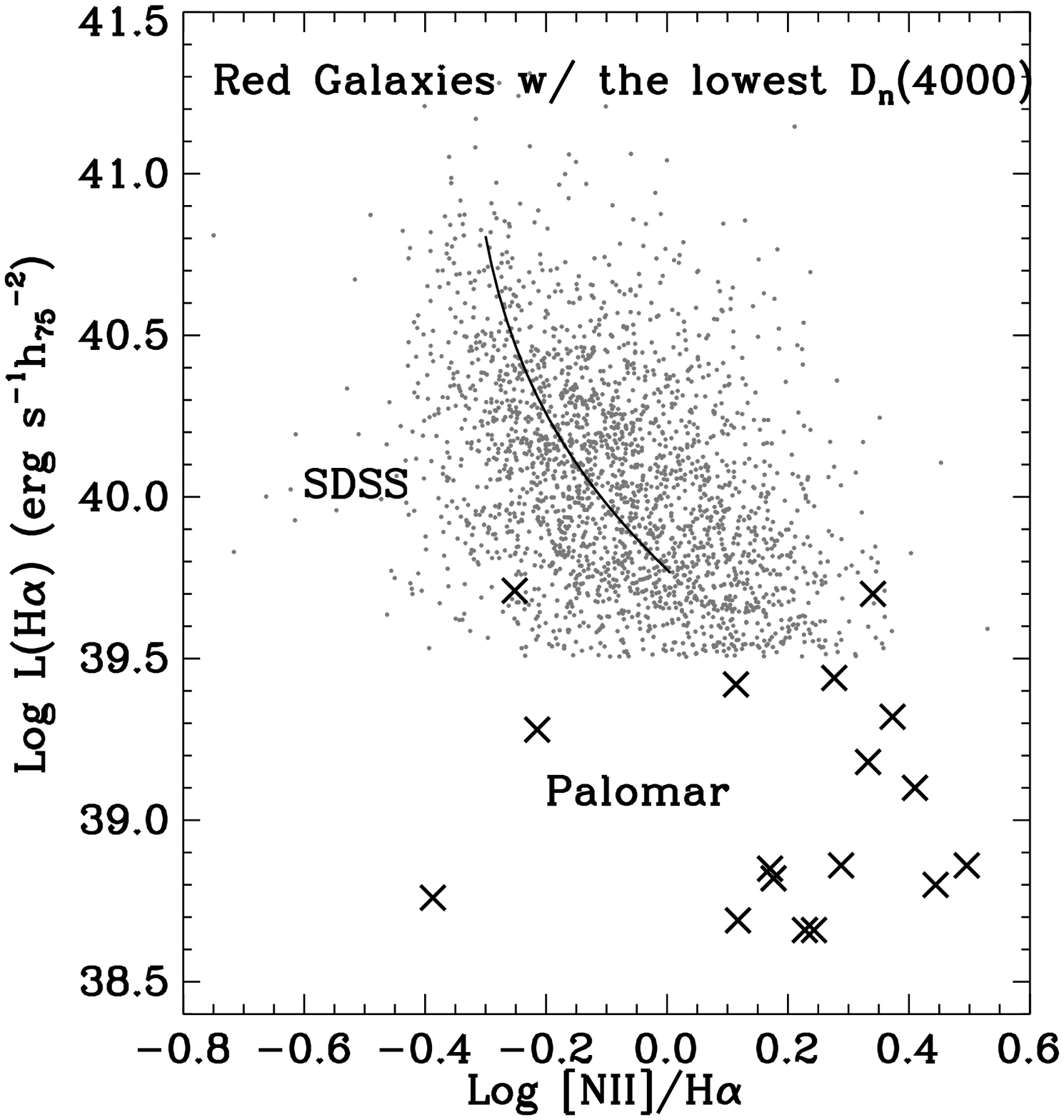}
\caption{Left: $\log L(\hal)$ vs. $\log \nii/\hal$ for passive red
  galaxies in the Palomar sample (dark crosses) and an SDSS comparison
  sample at $z\sim0.1$ (gray points). Only the brighter half (in
  \hal\ luminosity) of each sample is plotted, since the fainter half
  mostly have \hal\ undetected. The Palomar data is measured from
  nuclear aperture spectra while the SDSS data is measured from
  spectra integrated on much larger scales (5 $h_{75}^{-1}$ kpc
  diameter).  The curve represents the track followed by adding star
  formation to a typical Palomar nucleus. Clearly, star formation is
  not the cause of the luminosity difference.  The much larger
  \hal\ luminosities shown by the SDSS sample suggest that the line
  emission in them is spatially extended.  The right panel shows the
  red galaxies that we have excluded from the left panel because of
  possible star formation contamination. For the SDSS, this means
  those red galaxies with the lowest $D_n(4000)$; for Palomar, this
  means those red galaxies with Hubble types later than S0 and those
  classified as HII nuclei by \cite{HoFS97III}. The curve represents
  the track followed by adding star formation to a typical passive red
  galaxy in SDSS from the left panel.  For the SDSS samples in both
  panels, we also excluded a small fraction (11\% of the bright half
  in the left panel, and 1.5\% in the right panel) of galaxies with
  poorly measured \nii/\hal\ ratios (with uncertainties larger than
  0.25 dex), which has no practical impact on our conclusions.}
\label{fig:n2ha_hal_sloanpalomar}
\end{center}
\end{figure*}

In the left panel of Figure~\ref{fig:n2ha_hal_sloanpalomar}, we
compare $L(\hal)$ and \nii/\hal\ between the Palomar red sample and
the SDSS red sample at $z\sim0.1$. Not all galaxies in either sample
have \hal\ detected (64.4\% of the Palomar red sample and 52.0\% in
the SDSS red sample at $z\sim0.1$ have \hal\ detection). Therefore, we
only compare the brightest half of each volume-limited sample in \hal\ luminosity.

The SDSS red sample shows much brighter \hal\ luminosities and
slightly lower \nii/\hal\ ratios than the Palomar sample. Since 
both samples are volume-limited to the same magnitude cut, we
adopted the same sample selection, and any evolution effect over a
redshift difference of 0.1 should be tiny, this difference in \hal\ 
luminosities must be
caused mostly by the difference in the physical aperture size between
the Palomar survey and SDSS. The Palomar sample reflects the nuclear
properties of red galaxies while the SDSS sample reflects their
integrated properties on much larger scales.

The brightest 25th percentile in \hal\ luminosity for the $z\sim0.1$
SDSS sample (including non-detections) is $5.82\times10^{39} {\rm
  erg~s^{-1}~h_{75}^{-2}}$, nearly 7 times larger than that in the
Palomar sample ($0.85\times10^{39} {\rm erg~s^{-1}~h_{75}^{-2}}$). In
fact, even the median \hal\ emitter in the SDSS red sample is brighter
than the majority of nuclear \hal\ emitters in the Palomar red sample.
Nuclear emission in red galaxies is therefore only rarely as luminous
as found in the SDSS galaxies. Furthermore, we expect no strong
evolution in AGN activity between $z\sim0.1$ and $z\sim0$.  Thus, in
the SDSS galaxies, a substantial contribution to emission must come
from outside the nucleus, and therefore the \hal\ emission observed by SDSS 
in these $z\sim0.1$ passive red galaxies has to be {\it spatially extended}.

One might wonder if the spatially extended emission found in large
aperture measurements is due to low-level star formation in these
galaxies. We can simulate the expected \nii/\hal\ ratio and $L(\hal)$
by adding a typical star-forming emission-line spectrum to a typical
nuclear spectrum in the Palomar sample. We use the median
\nii/\hal\ and median \hal\ luminosity in the brighter half (in \hal)
of the Palomar red sample to represent a typical nucleus. For star
formation, we adopt an \nii/\hal\ ratio of 0.45, typical of a high
metallicity star-forming galaxy, which yields a conservatively high
\nii/\hal\ ratio. The result is shown by the curve in the left panel
of Fig.~\ref{fig:n2ha_hal_sloanpalomar}.  From the bottom end of the
curve to the top, the \hal\ luminosity contributed by the star
formation goes from 0 to 100 times that of the typical nucleus. The
curve misses the majority of the red galaxies at $z\sim0.1$.  Clearly,
the spatially extended line emission in red galaxies we selected at
$z\sim0.1$ is not powered by star formation.

As described above, in constructing this passively-evolving red galaxy
sample, we have removed potential star-forming contaminants by
removing the 30\% galaxies with the lowest $D_n(4000)$.  The right
panel of Figure~\ref{fig:n2ha_hal_sloanpalomar} shows the effect of
this procedure, where we plot the low $D_n(4000)$ galaxies plus those
Palomar red galaxies with Hubble types later than S0 for
comparison. In the SDSS, those red galaxies we have removed generally
have higher \hal\ luminosities and lower \nii/\hal\ than those we have
kept do. For the right panel, the curve indicates the track traced by
adding star formation to a typical passive red galaxy in the SDSS
sample from the left panel. From the bottom end of the curve to the
top, the \hal\ luminosity contributed by star formation goes from 0 to
10 times that of the median \hal\ luminosity in passive red galaxies
at this redshift. The curve traces the distribution fairly well,
suggesting that the line emission in red galaxies with low $D_n(4000)$
have more ongoing, low-level star formation than those red galaxies
with high $D_n(4000)$.

The right panel of Fig.~\ref{fig:n2ha_hal_sloanpalomar} also
demonstrates that the Palomar red galaxies we have removed from the
sample have fairly similar line luminosities and line ratios to those
Palomar red galaxies we have kept (except for the three HII nuclei,
with the lowest \nii/\hal\ ratios). This result reflects the fact that
the Palomar spectra have smaller physical apertures. Therefore,
although the criteria we used to remove star-forming contaminants
differs slightly for the Palomar sample than for the SDSS sample, the Palomar
sample is insensitive to these differences.

Now, we have shown that the line emission regions in passive red
galaxies in SDSS have to be spatially extended, simply because few
nuclear regions in red galaxies are luminous enough to explain the SDSS
results. The next question is which galaxies host these extended
\hal\ emission regions at $z\sim0$. Are they the same galaxies that
host those nuclear emission regions? The answer must be ``yes.''
Suppose they were not the same galaxies: then galaxies hosting these
extended \hal\ emission regions would need to have undetectable line
emission in their centers. No such galaxies are found by the SAURON
survey. As shown by \cite{Sarzi06}, in a representative sample of 48
early-type galaxies in the nearby universe, all galaxies with emission
lines detectable have their line flux peaking at the center, and the
distribution is nearly always extended. Therefore, we conclude that
most, if not all, red galaxies that have nuclear emission line regions
also have extended line-emitting regions, and vice versa. The host
galaxies of nuclear line emitting regions and those of extended line
emitting regions are largely the same population.

\subsection{Emission line surface brightness profile}\label{sec:surfacebrightness}

In this section, we compare the emission line luminosity distributions
among passive red galaxies at a series of redshifts, which translates
to a series of apertures, to investigate the average surface
brightness profile of their line emission. We bin the SDSS sample with
$0.01<z<0.1$ into 9 redshift bins with a binsize of 0.01. We limit the
Palomar sample to only those galaxies at $D<40{\rm Mpc}$, which
corresponds to $z=0.01$, for the lowest redshift bin.

\begin{figure}
\begin{center}
\includegraphics[totalheight=0.35\textheight]{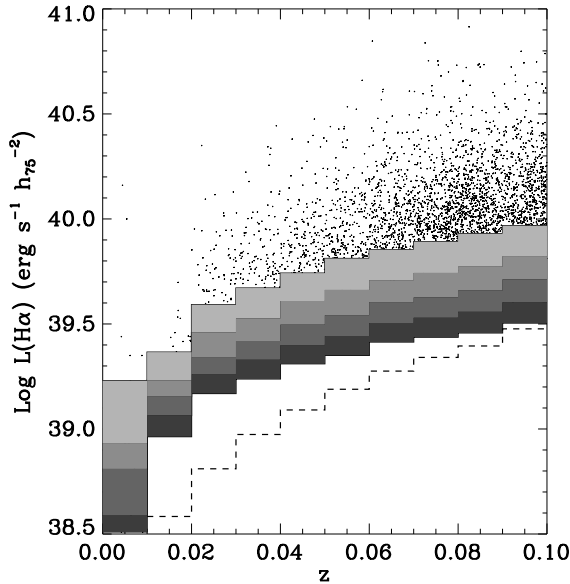}
\caption{\hal\ luminosity distributions of passive red galaxies as a
function of redshifts. The first bin at $z<0.01$ is from the Palomar sample and
the rest are from the SDSS sample. Only the brighter half of the sample in each
redshift bin is plotted. The brightest 10 percent of galaxies in \hal\
in each bin are plotted as points. The gray scales indicate the ranges
populated by different percentiles in each bin: brightest 10-20th,
20-30th, 30-40th, and 40-50th, from top to bottom, respectively. The dashed
line at the bottom indicates the 3$\sigma$ detection limit in SDSS,
which is 3 times the luminosity corresponding to the median \hal\ flux error in each bin.
Clearly, the \hal\ luminosity increases with increasing redshift or
aperture size.}
\label{fig:halumdist_all}
\end{center}
\end{figure}

\begin{figure}
\begin{center}
\includegraphics[totalheight=0.35\textheight]{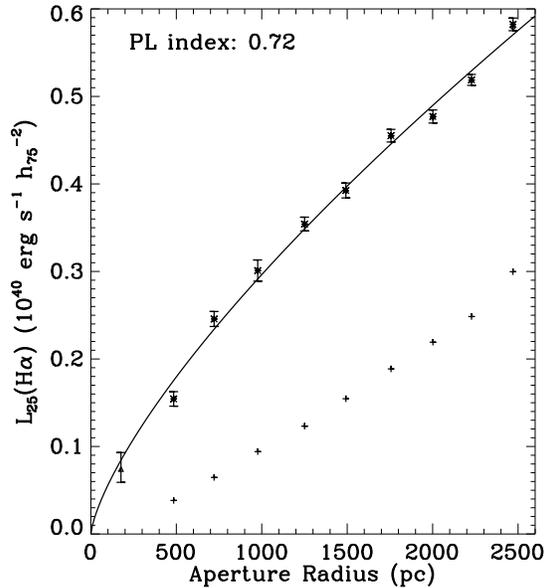}
\caption{The 25th percentile (counting from the brightest)
  \hal\ luminosity (asterisks with error bars) measured with SDSS fibers in 
  a sample of
  non-star-forming red sequence galaxies as a function of the physical
  scale probed by the fiber. The triangle point represents the
  measurement in the Palomar sample, which is fairly consistent with
  the extrapolation of the power-law fit on small scales. The
  uncertainties of these measurements are measured with bootstrap
  resampling. The '+' signs at the bottom indicate the 3$\sigma$ detection 
  limits.}
\label{fig:halum25_z}
\end{center}
\end{figure}

Figure~\ref{fig:halumdist_all} shows the \hal\ luminosity distribution
as a function of redshift. In each bin, we only plot the brighter half
of the sample in \hal.
With increasing redshift, i.e., increasing aperture, the
\hal\ luminosities increase. Thus, the \hal\ luminosities observed
with larger apertures have to have significant contributions from
spatially extended emission line regions.  In each redshift bin, we
sort all passive red galaxies by their \hal\ luminosity.
Figure~\ref{fig:halum25_z} plots the brightest 25th percentile
\hal\ luminosity as a function of physical scale covered by the SDSS
fiber. The 25th percentile is safely above the detection threshold at
all redshifts. The luminosity increases with scale roughly as a power
law with an index of $0.72$, as shown by the power-law fit in
Fig.~\ref{fig:halum25_z}. As this is the integrated luminosity within
radius $r$, it indicates the average surface brightness profile
follows $r^{-1.28}$. This slope is fairly consistent with what
\cite{Sarzi10} found in nearby early-type galaxies targeted by the
SAURON survey (see their Figure 4). We also show the same measurement
of the Palomar sample (with late-type galaxies removed) at the median
effective radius probed by the $2\arcsec\times4\arcsec$ aperture in
the Palomar survey, which we treat as equivalent to a circular
aperture with 3\arcsec diameter.  The 25th percentile in the Palomar
sample is fairly consistent with the power-law fit to the SDSS data
points.  This evidence further strengthens the conclusion that the
nuclear and the extended line emitting regions exist in the same
galaxy population.

However, the extended line emission alone does not constrain the
source of the ionizing radiation. Contrary to intuition, the extended
emission could also be produced by a central ionizing source, such as
an AGN. The emission line brightness profile depends on many factors:
the ionizing flux profile, the density profile, the gas filling
factor, and how the typical size of the gas clouds change with radius.
We leave the detailed calculations to \S\ref{sec:sbprofile}.

\section{Line Ratio Gradient}

With the above technique, we can also check if the line ratio
distribution in this population changes between
redshifts/apertures. This check can only be done on those galaxies
with detectable line emission.
To ensure low uncertainty on the line ratio measurement, we choose
only the brightest 25\% in total emission line luminosity at each
redshift and compare their various line ratios. To avoid bias on the
line ratios, instead of selecting the brightest 25\% in
\hal\ luminosity, we base the selection on the total luminosity of the
several strongest emission lines available in the spectra
(\hal+\niibw+\oiiibw+\siiw). This combination is a better proxy of the
total line emission output than L(\hal).

\begin{figure*}
\begin{center}
\includegraphics[totalheight=0.7\textheight, angle=90,viewport=0 0 470 770,clip]{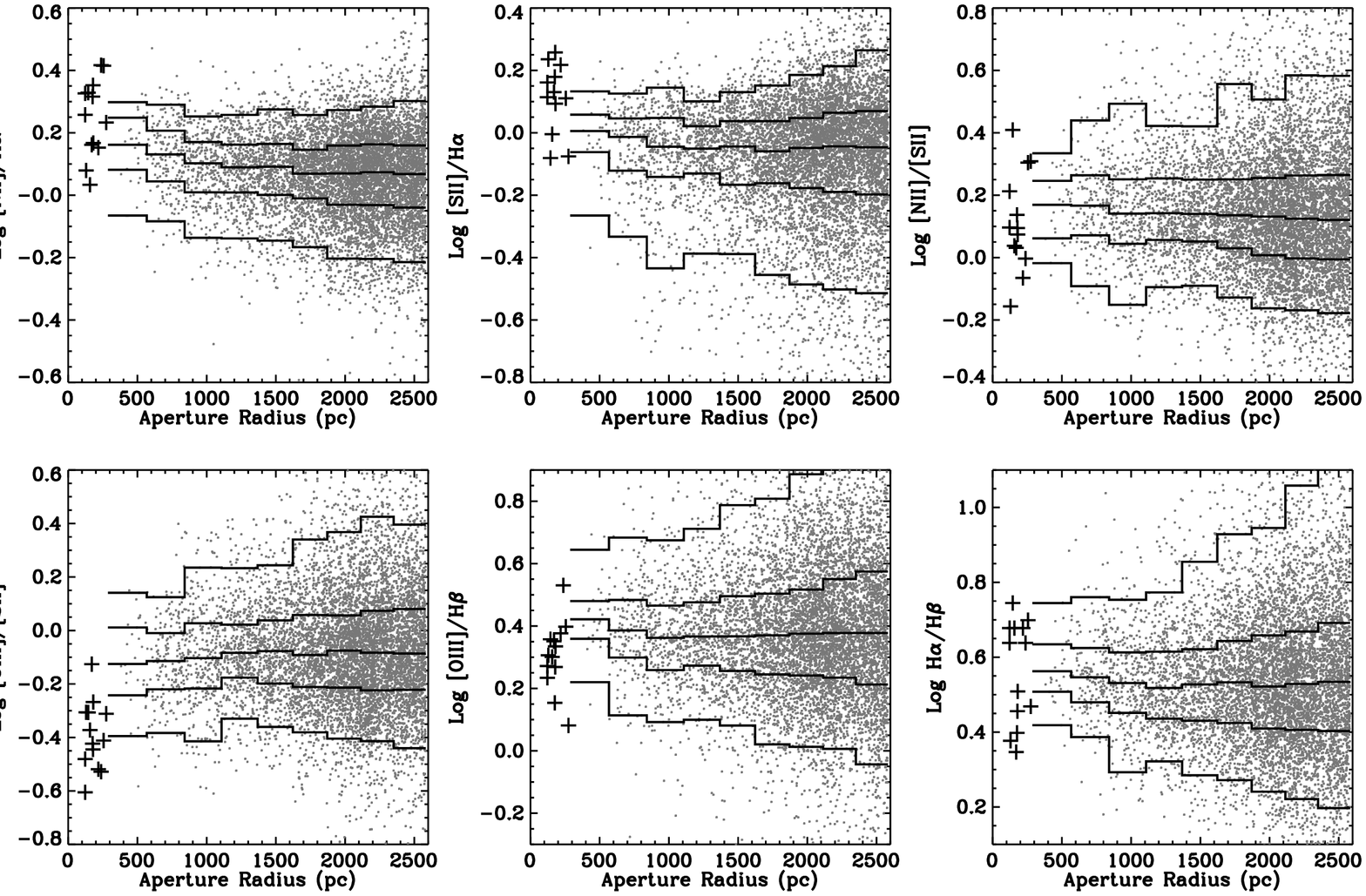}
\caption{Integrated line ratio distribution as a function of the aperture radius for passive red galaxies in SDSS (gray points) and the Palomar sample (red crosses). The line ratios are "integrated" in the sense that they are derived using the fluxes interior to the indicated aperture radii. 
Only the 25\% brightest (in \hal) galaxies in each redshift bin are plotted to ensure small uncertainty on the line ratio measurement. The lines indicate the 5-, 25-,50-,75-, 95-percentile points in each redshift bin. The aperture of Palomar galaxies are computed using an effective radius of 1.5". }
\label{fig:lineratio_all}
\end{center}
\end{figure*}

\begin{figure*}
\begin{center}
\includegraphics[totalheight=0.7\textheight, angle=90,viewport=0 0 470 770,clip]{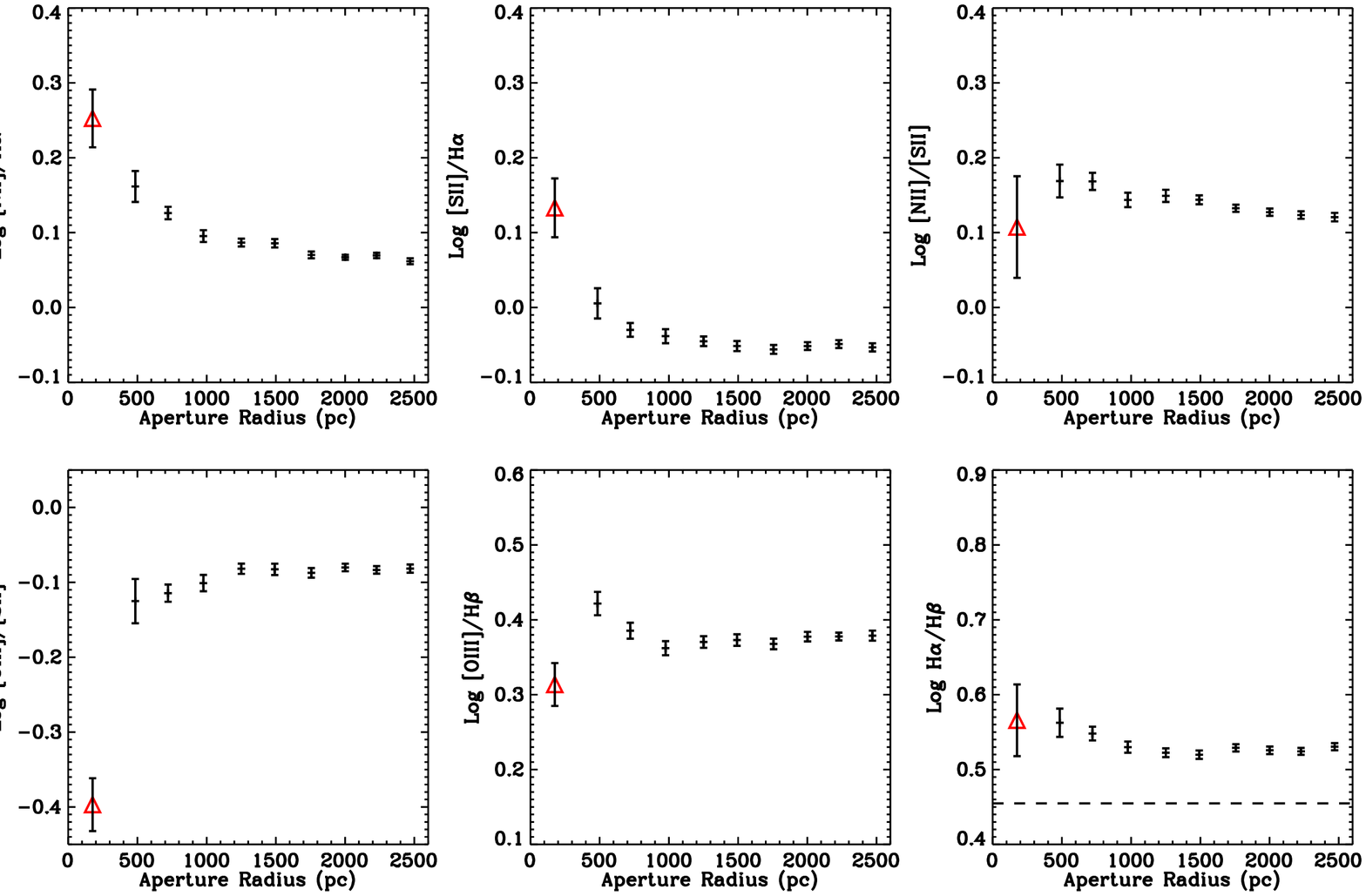}
\caption{Median integrated line ratios as a function of aperture radius for the
  25\% brightest line emitters among the red galaxies in SDSS
  ($>300~{\rm pc}$) and the Palomar sample ($<300~{\rm pc}$). 
  For the median, we use the biweight center estimator \citep{Beers90}, 
  which is more robust than the median for small samples. 
  The uncertainty is given by the jackknife error of the biweight estimator.
  All panels cover 0.5
  dex in y-axis for the ease of comparison. The horizontal line in
  \hal/\hb\ panel marks the Case B ratio of 2.85.}
\label{fig:lineratio_scale}
\end{center}
\end{figure*}

Figure~\ref{fig:lineratio_all} shows the full distribution of several
line ratios as a function of aperture radius. We also plot those
galaxies in the Palomar red sample with $D<40~{\rm Mpc}$ to probe the
smallest scales. As for the SDSS sample, we only select the brightest
25\% galaxies in total observed emission line luminosity.

Fig.~\ref{fig:lineratio_all} shows the distribution of line ratios in
the red galaxies. At small aperture radii, the scatter is dominated by
intrinsic variations in line ratios among galaxies. At the large
aperture end, the uncertainty in line measurements starts to dominate
the scatter, leading to a slight increase in the width of the
distribution. In most line ratios, the intrinsic variation has a
standard deviation of approximately 0.1 dex.

Interestingly, in some line ratios, the median of the distribution
changes systematically with aperture radius, most notably in
\nii/\hal, \sii/\hal, and \oiii/\sii. The trends also extend to the
Palomar sample at the smallest scales. The changes are so large in
\sii/\hal\ and \oiii/\sii ratios that most of the Palomar sample
populates only one side of the median of SDSS sample even in its first
bin.

Figure~\ref{fig:lineratio_scale} shows how the median line ratios
change with aperture size, with the error bars giving the
uncertainties of the median estimates. The systematic changes in
\nii/\hal, \sii/\hal, and \oiii/\sii\ with radius are significant and
the Palomar sample confirms the trend on small scales.

One might worry that the change in \oiii/\sii\ ratio could be due to
dust extinction, because these two lines are separated in
wavelength. However, the \hal/\hb\ ratio is nearly constant with
radius, indicating that the average dust extinction does not vary
much. The values of the median \hal/\hb\ ratio are also close to the
Case B prediction of 2.85 (or 3.1 if collisional excitation is
included), indicating that the level of dust extinction is
low. Applying extinction corrections to each galaxy in the sample only
shifts the median \oiii/\sii\ ratio up by a nearly constant $\sim0.1$
dex in each bin.

The line ratios presented in Fig.~\ref{fig:lineratio_scale} are
cumulative measurements: they reflect the luminosity-weighted average line ratios within aperture radius $r$, rather than that in an annulus at radius $r$. 
We need to combine these median line ratios in integrated apertures with the median 
line luminosity of this sample in corresponding apertures, to derive the average line ratios in circular annuli at radius $r$. 

\begin{figure}
\begin{center}
\includegraphics[totalheight=0.35\textheight]{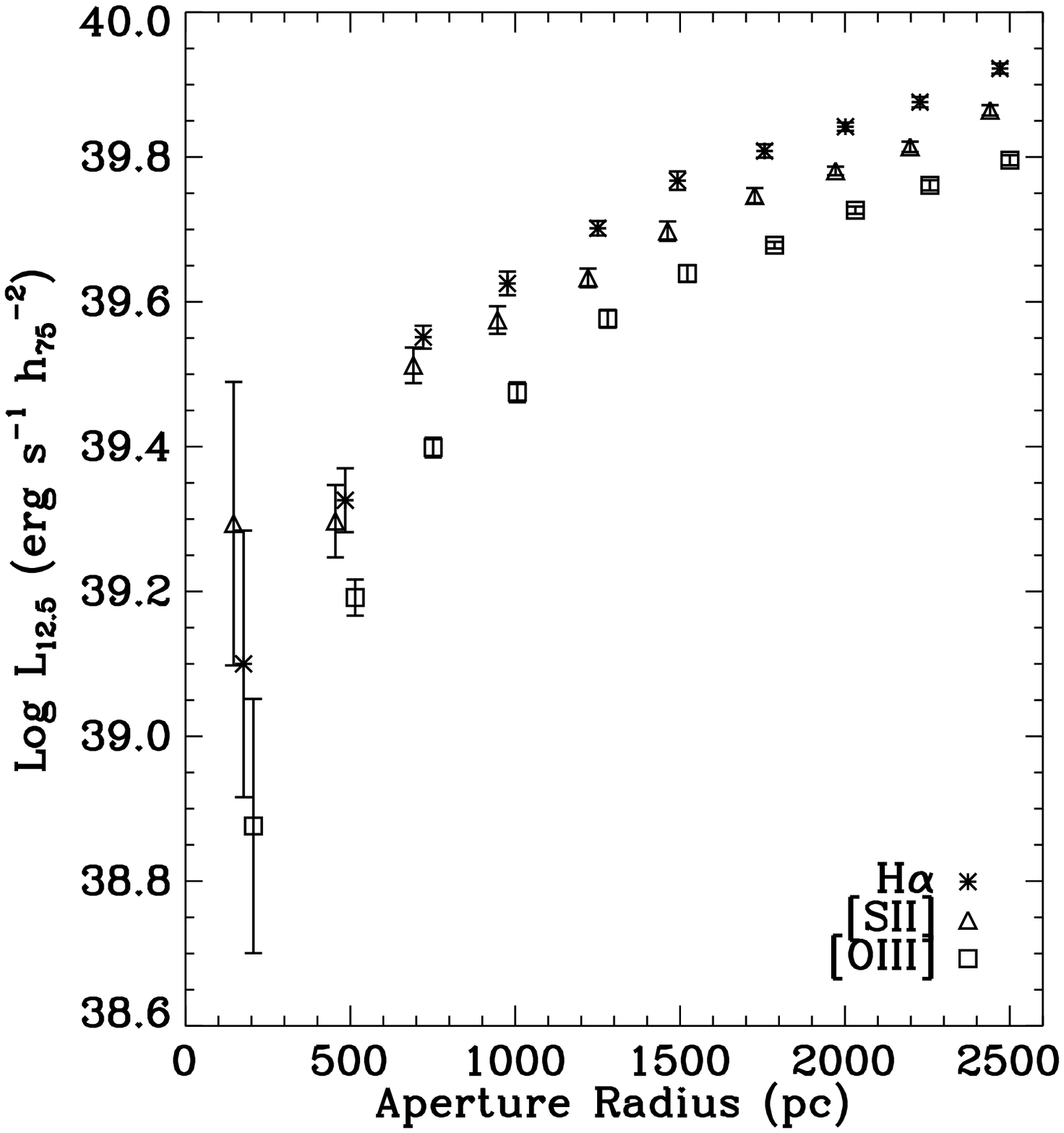}
\caption{The median integrated \hal\ (asterisk), \sii\ (triangle), and \oiii\ (square) luminosities as a function of 
aperture among the 25\% passive red galaxies (in each bin) with the brightest total emission line luminosity. The smallest scale is probed by the Palomar sample and the larger scales are probed by the SDSS sample. The \sii\ and \oiii\ points are slightly shifted in the horizontal direction for clarity. The uncertainties of these measurements are measured with bootstrap resampling.}
\label{fig:linelum12_z}
\end{center}
\end{figure}

In Fig.~\ref{fig:linelum12_z}, we show the median line luminosities of
this sample in corresponding apertures for \hal, \sii, and \oiii. This
plot is similar to Fig.~\ref{fig:halum25_z}, but uses only the 25\%
brightest galaxies in total emission line luminosity. From this
figure, it is evident that the \sii\ line has a very different surface
brightness profile from \hal\ and \oiii, consistent with the trends in
integrated line ratios.

\begin{figure*}
\begin{center}
\includegraphics[totalheight=0.7\textheight, angle=90,viewport=0 0 470 770,clip]{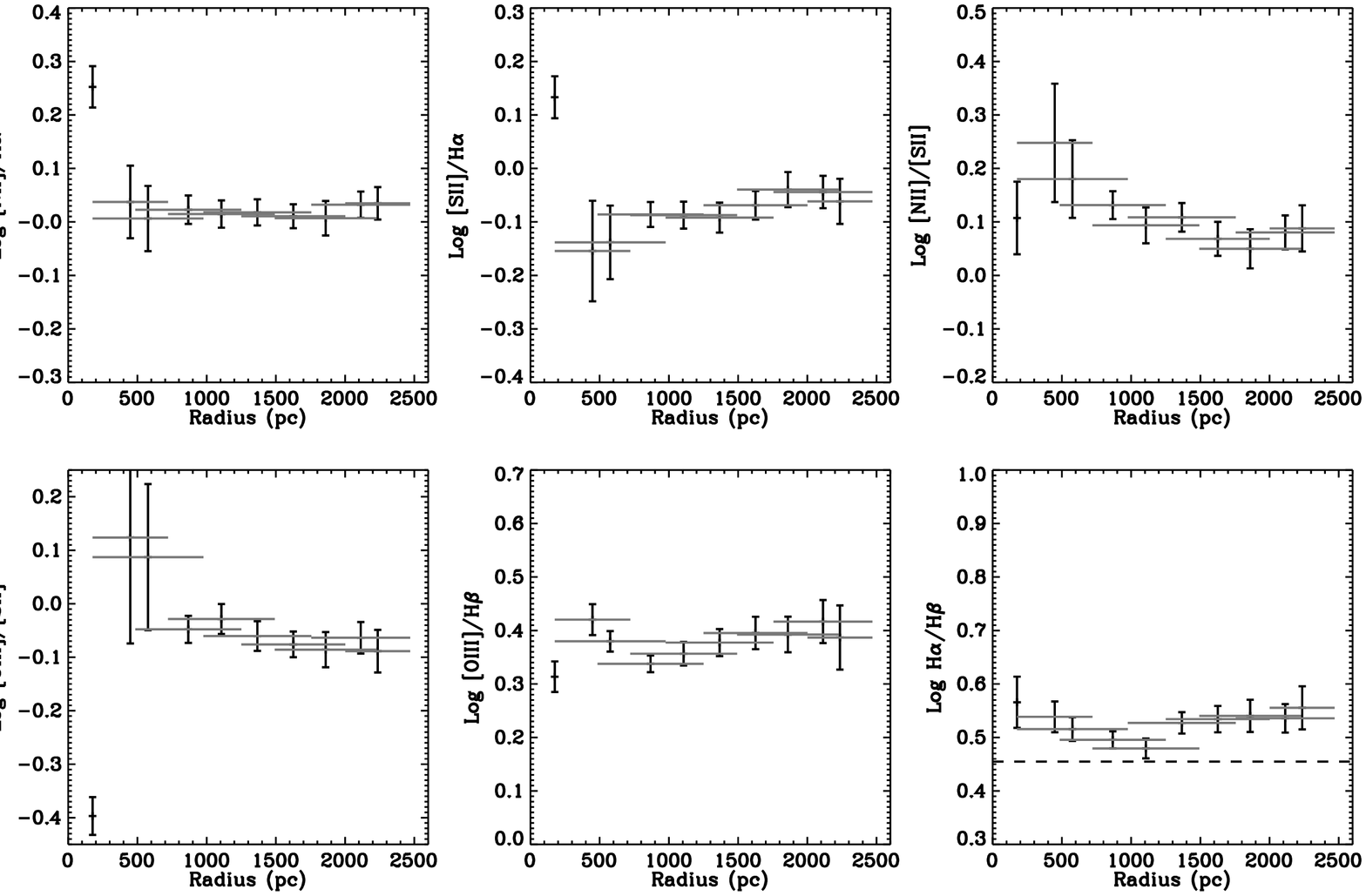}
\caption{Average line ratios in annuli as a function of radius for passive red galaxies. This is derived by combining the integrated line ratios in Fig.~\ref{fig:lineratio_scale} with the integrated luminosities in Fig.~\ref{fig:linelum12_z}, through equations similar to Eqn.~\ref{eqn:difflineratio}.  The horizontal bar of each data point indicates the range of radius covered by the annulus. 
Note the y-axis in each panel covers a larger range (0.7 dex) than Fig.~\ref{fig:lineratio_scale}. 
}
\label{fig:lineratio_ring_data}
\end{center}
\end{figure*}

With the integrated line ratios and integrated luminosities, we can compute the line ratio in annuli. For example, the average \nii/\hal\ ratio between radius $r_i$ and $r_j$ can be computed by the following equations,
\begin{equation}
\left\langle{\nii \over \hal}\right\rangle_{r_i<r<r_j} =
            {\left\langle{\nii \over
                \hal}\right\rangle_{r<r_j}L_j(\hal) -
              \left\langle{\nii\over\hal}\right\rangle_{r<r_i}L_i(\hal) \over
                          L_j(\hal)-L_i(\hal) },
\label{eqn:difflineratio}			  
\end{equation}
where $L_i(\hal)$ and $L_j(\hal)$ are the median luminosities in apertures with radius $r_i$ and $r_j$, respectively. The \sii/\hal\ and \hal/\hb\ ratios in annuli are also computed by combining integrated line ratios with the median \hal\ luminosities. The \nii/\sii\ and \oiii/\sii\ ratios are computed by combining integrated line ratios with the median \sii\ luminosities. The \oiii/\hb\ ratios are computed by combining with the median \oiii\ luminosities. 
Calculating this between every consecutive bin results in large uncertainties, due to the large fractional error in the luminosity differences. 
We therefore calculate the annulus line ratios using aperture pairs: $[i,j] = [1,3],[1,4],[2,5],[3,6],[4,7],[5,8],[6,9],[7,10], {\rm and} [8,10]$, where Aperture 1 is the Palomar aperture and Aperture 10 is the aperture for SDSS galaxies at $0.09<z<0.1$. 
The results are shown in Fig.~\ref{fig:lineratio_ring_data}

Although the uncertainties become much larger, the differences between
the innermost bin and the outer bins remain robust in \nii/\hal,
\sii/\hal, and \oiii/\sii. The second bin in each panel always shows a
very large uncertainty. This is due to the dramatic change in line
ratios between the first few bins and their larger uncertainties.
In Table~\ref{tab:lineratio_change}, we list the line ratios of the
inner most bin and the median line ratios of the outer bins ($700<r<2500~{\rm pc}$). 

The line ratio gradients we observe have been in principle detectable in past long-slit spectroscopic surveys of early-type galaxies \cite[e.g.,][]{Phillips86, Kim89,Zeilinger96, HoFS97III, Caon00}. However, these authors either did not have data with sufficient quality or did not look at the line ratio gradient at all. The only exception is \cite{Zeilinger96}, who showed that the \nii/\hal\ ratio decreases outwards in four galaxies. However, like many of these past surveys, their data did not have wide enough wavelength coverage to cover multiple line ratios, which would be critical for identifying the cause of the line ratio gradients. Recently, \cite{Annibali10} investigated the line ratio gradients using long-slit spectra with wide wavelength coverage for a sample of relatively gas-rich early-type galaxies. They found that the \nii/\hal\ ratios in most of them decrease with radius, consistent with our result. However, they did not look for gradients in other line ratios, except for \oiii/\hb, which is very often too noisy to support firm conclusions. 

We will discuss in \S\ref{sec:discussion} what changes in physical conditions
are required to produce such changes in line ratios.

\begin{table}
\begin{center}
\caption{Median line ratios within 300~pc radius and outside 700~pc}
\begin{tabular}{lccc}
\hline\hline
Location & $\log$\nii/\hal & $\log$\sii/\hal & $\log$ \oiii/\sii \\\hline
Inner 300~pc & $0.25\pm0.03$ & $0.13\pm0.04$ & $-0.40\pm0.04$ \\
Outside 700~pc & $0.01\pm0.01$ & $-0.07\pm0.01$ & $-0.07\pm0.01$ \\
\hline
\end{tabular}
\label{tab:lineratio_change}
\end{center}
\end{table}

\section{Clues from Line Widths} \label{sec:linewidth}

We initally thought that the line ratio gradient would produce
different line widths in different emission lines, due to varying
kinematics of the line emitting clouds across each galaxy. If this
were so, it could provide a complementary constraint on the radial
distribution of the emission. The data do indeed show different line
widths for different lines. However, we have concluded that this
variation is probably not due to line ratio and kinematic gradients,
primarily because the line width differences are not a function of
aperture size.  We describe our investigation of line width
differences in this section.

Because the width measurement is noisy on SDSS spectra, we need a control sample to demonstrate that our line width measurement is robust. The star-forming galaxies provides such a comparison. In a pure star-forming galaxies, the line emission originates mostly from HII regions photoionized by O and B stars. The line width in integrated spectra reflects the rotation velocity of the galaxy. 
In the absence of strong metallicity gradient, we would observe approximately the same line ratios in all HII regions. In this case, all lines should display the same line width. 

\begin{figure*}
\begin{center}
\includegraphics[totalheight=0.35\textheight]{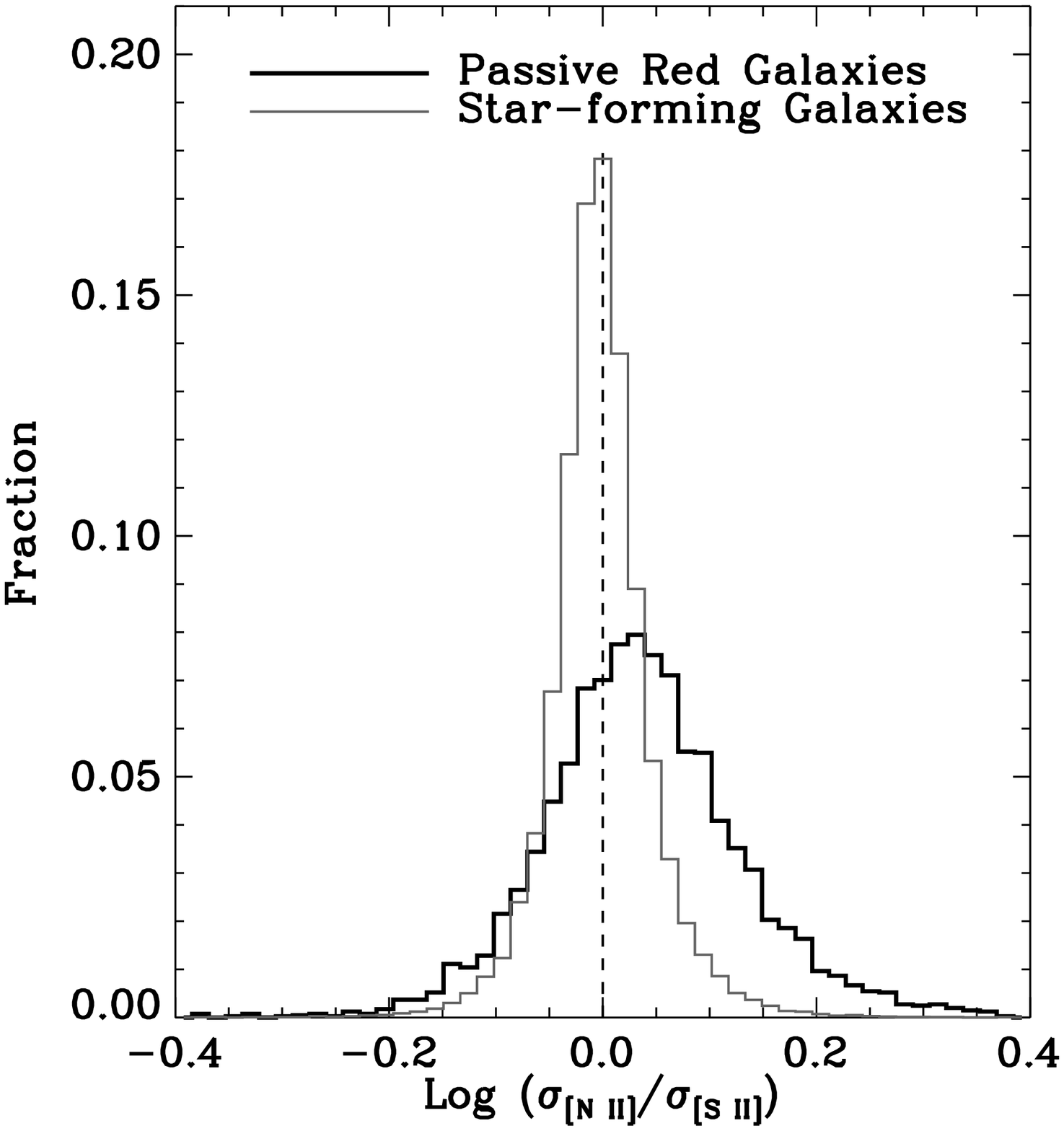}
\includegraphics[totalheight=0.35\textheight]{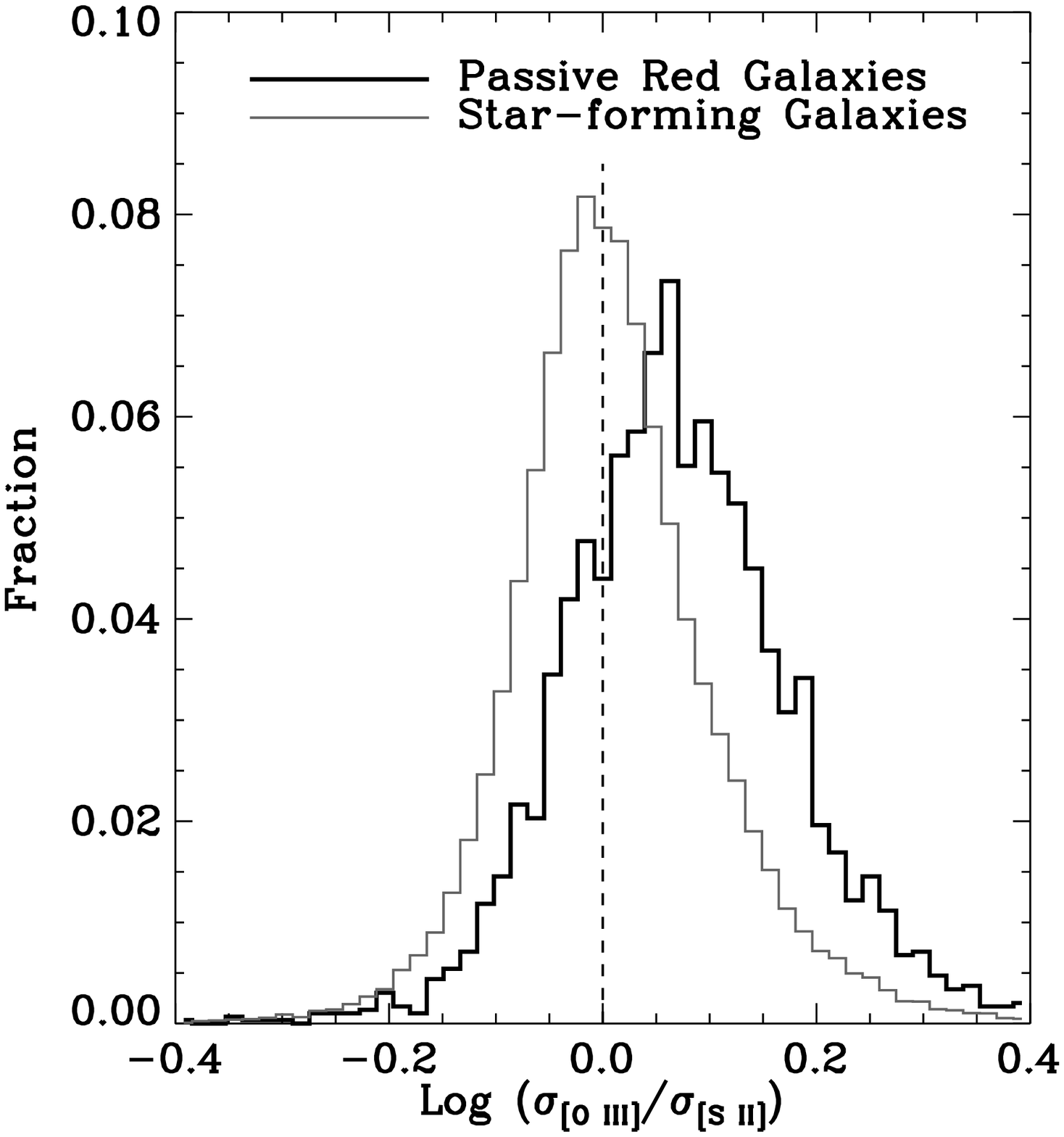}
\includegraphics[totalheight=0.35\textheight]{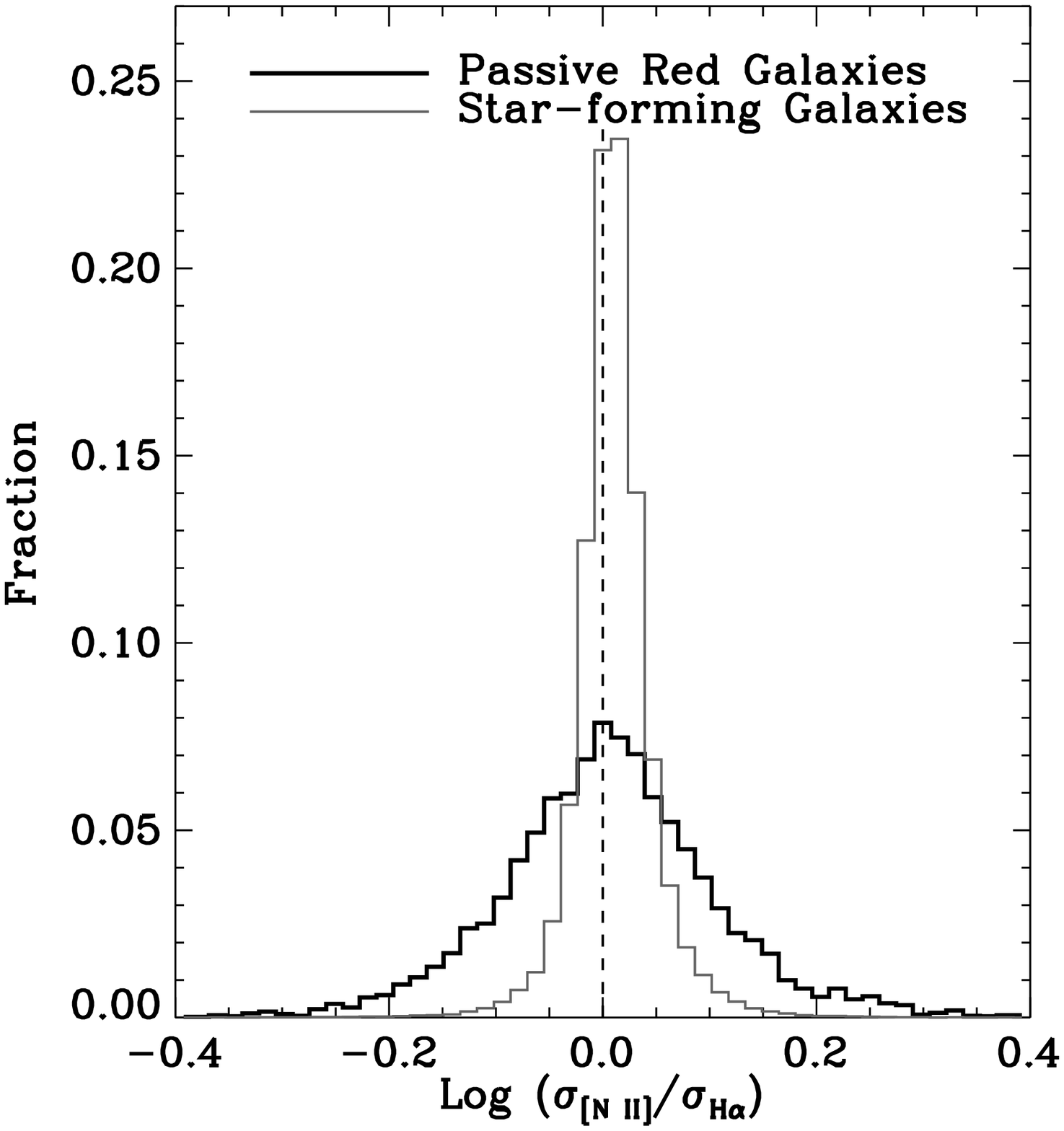}
\includegraphics[totalheight=0.35\textheight]{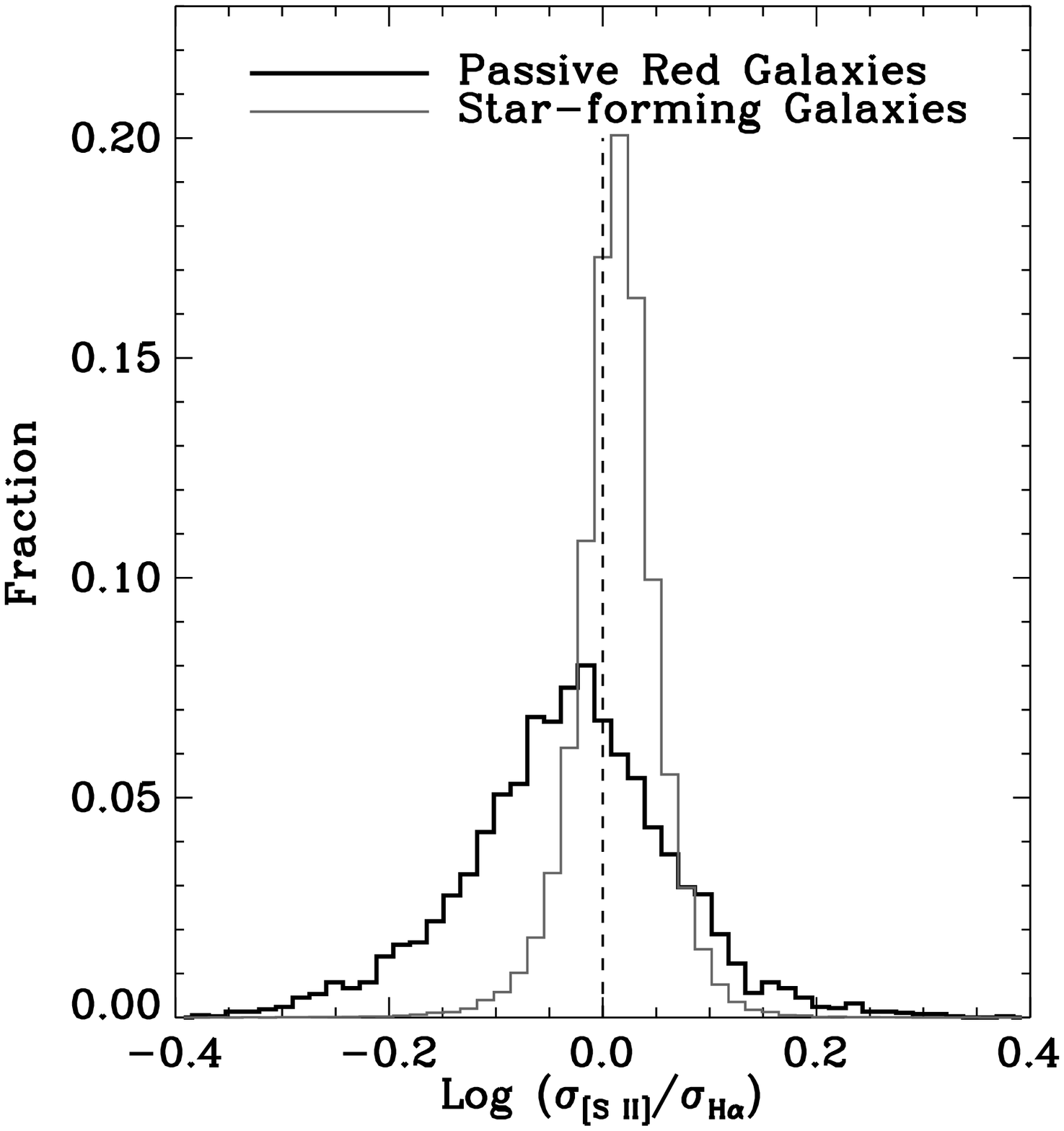}
\caption{Panel (a): Distributions of line width ratio between \nii\ and \sii\ for star-forming galaxies (thin-lined histogram ) and passive red galaxies (thick-lined histograms). \nii\ in passive red galaxies are systematically wider than \sii\ in both samples. Panel (b), (c), (d) show the same comparisons in \oiii-to-\sii\ width ratio, \nii-to-\hal\ width ratio, and \sii-to-\hal\ width ratio, respectively.}
\label{fig:widthratio}
\end{center}
\end{figure*}

We choose a star forming galaxy sample using the line ratio criteria
described by \cite{Kewley06}. Basically, the galaxies in this sample
are selected to fall in the star-forming branch on all three
diagnostic diagrams (\oiii/\hb\ vs. \nii/\hal, \sii/\hal, and
\oi/\hal). We compare this star-forming galaxy sample with the sample
we used for deriving the line ratio profile, namely the top 25\%
passive red galaxies that have the highest total emission line
luminosities in each redshift bin. We combine all the redshift bins
together. In addition, to ensure good measurements on the line width
ratio, we require the uncertainty of the line width ratio to be
smaller than 0.1 dex.

In Figure~\ref{fig:widthratio}, we plot the distributions of width
ratio in four line pairs for galaxies in the star-forming sample
(thin-lined histograms) and in the passive red galaxy sample (thick-lined
histograms). In all line pairs, the distribution of star-forming
galaxies always show a fairly symmetric and narrow distribution
peaking around zero in logarithmic space, meaning all the lines have
roughly the same width. This proves that our line width measurement is
robust. It also indicates that line ratios in star-forming HII regions
do not correlate strongly with the velocity of the HII regions.  The
\oiii-to-\sii\ pair may be an exception; for star-forming galaxies
this width ratio has a wider distribution than the other line pairs.
This broad distribution probably reflects intrinsic variation in the
population. It might be caused by variation in the disturbed component
of the diffuse ionized medium in star-forming galaxies, which produces
strong and wide \oiii\ lines \citep{Wang97}. We leave this question
for future investigations.

However, for line-emitting passive red galaxies, the distributions in
\nii-to-\sii, \oiii-to-\sii, \sii-to-\hal\ width ratios do not peak
around zero in log space. Their offsets from zero are highly
significant. On average, the \nii\ and \hal\ lines in them are wider
than \sii\ lines by 8\% and the \oiii\ lines are wider than \sii\ by
16\% (Table~\ref{tab:medianratio}).  
This means the \oiii/\sii\ ratios in the line wings are 
higher than that in the line center, and the higher velocity clouds 
which contribute to the wings must have higher \oiii/\sii\ ratios.
The same must be true for \nii/\sii\ and \hal/\sii\ ratios.
This indicates that in these
galaxies, line emitting clouds do not have uniform line ratios and the
line ratio must correlate with the velocity of the clouds.

In the previous section, we found the line ratios have a systematic variation 
with radius. Could these line ratio variations produce the line width differences?

Given the fact that \oiii/\sii\ ratio increases with radius, to obatin a wider \oiii\ line than \sii, the line-of-sight velocity broadening must also increase with radius. The broadening could either come from random motions or ordered rotation. As shown by previous long slit \citep{Phillips86,Kim89,Zeilinger96} and IFU \citep{Sarzi06} observations, in most early-type galaxies, the kinematics of the gas is largely consistent with disk-like rotation with rotation velocity increasing outwards (though see \citealt{HeckmanBvB89} for some exceptions). This is consistent with the requirement here. The \nii-to-\sii\ and \sii-to-\hal\ width ratios could also be consistent with their respective line ratio gradients (if the decrease in \nii/\sii\ towards the center is real, as indicated by the Palomar data point). 
However, the \nii\ show the same width as \hal, inconsistent with the expectation. 

\begin{figure}
\begin{center}
\includegraphics[totalheight=0.35\textheight]{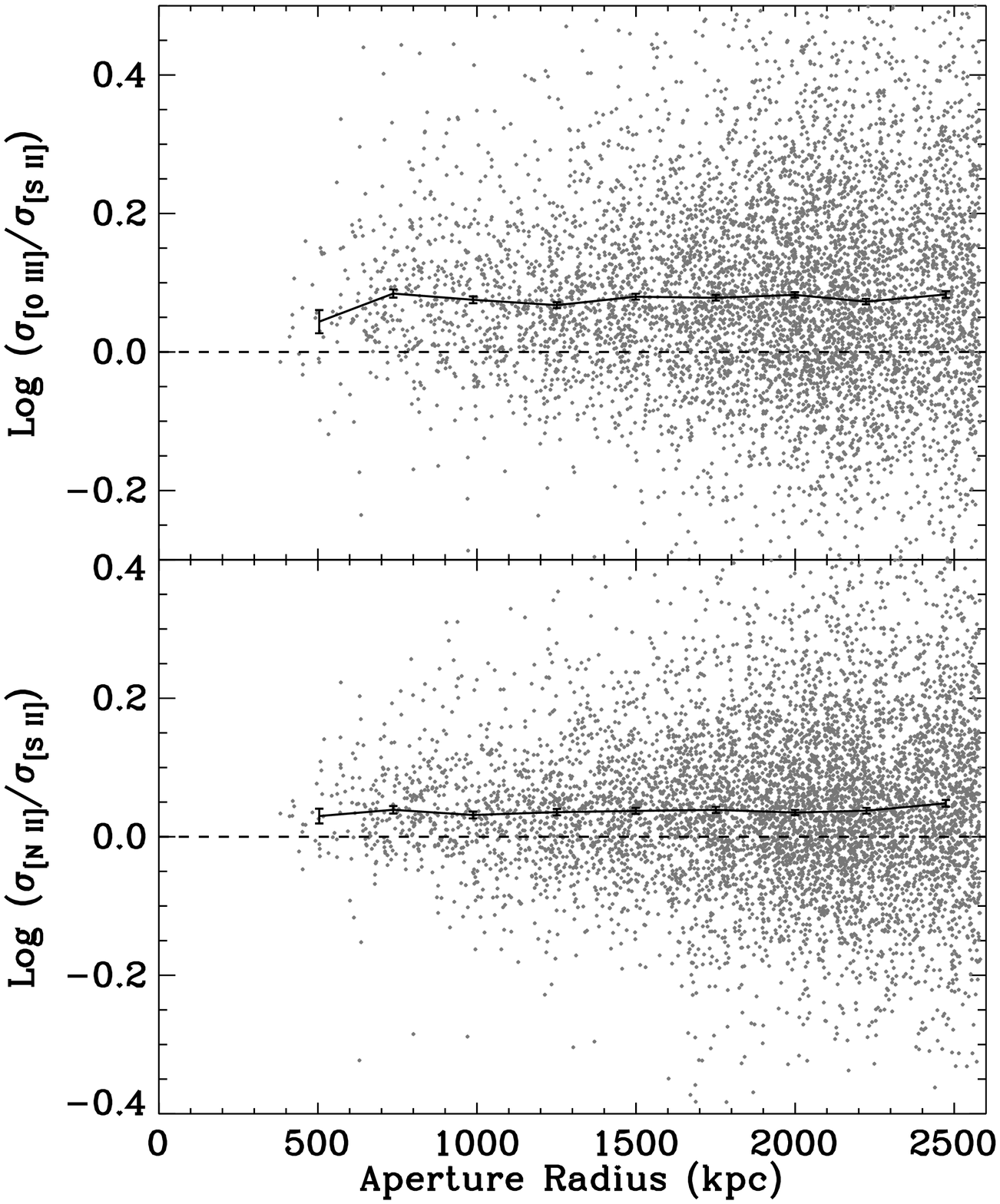}
\caption{Top panel: the \oiii-to-\sii\ width ratio distribution as a function of aperture radius for the 25\% passive red galaxies with the highest total line luminosity. The dark points with error bars indicate the median width ratio in each redshift bin. Bottom panel: same plot for the width ratio between \nii\ and \sii.}
\label{fig:lw_scale}
\end{center}
\end{figure}

On the other hand, if the width difference is indeed caused by the line ratio gradient and the rotation of the gas disk, the line width difference should gets smaller in larger apertures, since we expect the flux to be increasingly dominated by the outskirts where the line ratio profile gets flat. Figure~\ref{fig:lw_scale} shows the line width ratio distributions as a function of aperture radius. Apparently, the average line width ratios are roughly constant, independent of the aperture size. This is inconsistent with the expectation of the model.

We demonstrate this inconsistency quantitatively by simulating the
expected line width ratios using a toy model of a rotating gas disk in
a spherically-symmetric galaxy with stellar mass of
$7.5\times10^{10}M_\odot$ and an effective radius of 4.8 kpc, the
median values among our 25\% passive red galaxy sample with the brightest total line 
luminosity. The rotation curve
of the gas disk is set by the stellar density profile, which is
assumed to be a $\gamma$-model described by \cite{Dehnen93} with
$\gamma=1.5$. This stellar density profile gives a stellar surface
brightness profile closely resembling the de Vaucouleurs' $R^{1/4}$
profile. The integrated luminosity profile in \oiii\ is fixed to be a
power law with index of 0.77, measured by fitting the data points in
Fig.~\ref{fig:linelum12_z}.  The inclination of the disk is set at
$60^{\circ}$. We assume that the line ratios at each point in the disk
are solely dependent on radius. We model the logarithm of the line 
ratio profile as a broken power-law of the form,
\begin{equation}
\log {\oiii\ \over \sii} = \left\{ \begin{array}{rl}
     A (r/r_0)^{\gamma_1}-0.5 &, r < r_0 \\
     (A-0.4) (r/r_0)^{\gamma_2}-0.1 &, r \ge r_0 
     \end{array} \right.
\end{equation}
Based on the trend seen in Fig.~\ref{fig:lineratio_ring_data}, 
we fix the model to have $\log \oiii/\sii$ equal to -0.5 (an arbitrary 
choice) at $r=0$ and asymptote to -0.1 as $r \rightarrow \infty$. 
We fit the model to the integrated line ratios rather than the
differentiated line ratios, since the former have independent
uncertainties.

The velocity dispersion in the disk is assumed to be a constant
everywhere and equal to $50 {\rm km s^{-1}}$. %one-third of the maximum rotation velocity. 
Many gas kinematic studies have shown that the velocity
dispersion is likely to increase towards the center. Here we assume
the extreme case of flat dispersion, since using an increasing
velocity dispersion towards the center would erase the line width
differences.

We employ a Markov-Chain Monte Carlo technique to find a large number 
of models that best fit the data. Among the 10 data points, we ignored
the second bin ($r\sim500~{\rm pc}$) in the fitting, as including it 
makes the fit difficult. We chose five typical but different models to 
illustrate the trend expected in line width differences as 
aperture radius increases. The model parameters are given in 
Table.~\ref{tab:modelpara}.

\begin{table}
\begin{center}
\caption{Line-ratio Profile Model Parameters}
\begin{tabular}{c|c|c|c}
A & $r_0$ (kpc) & $\gamma_1$ & $\gamma_2$ \\ \hline
1.86 & 345.9 & 3.69 &  -2.44\\
1.56 & 309.2 & 3.82 & -2.04 \\
1.21 & 402.3 & 2.29 & -1.98 \\
0.982 & 565.1 & 1.30 & -2.30 \\
0.868 & 364.7 & 1.58 & -1.46 \\
\end{tabular}
\label{tab:modelpara}
\end{center}
\end{table}

\begin{figure*}
\begin{center}
\includegraphics[totalheight=0.35\textheight]{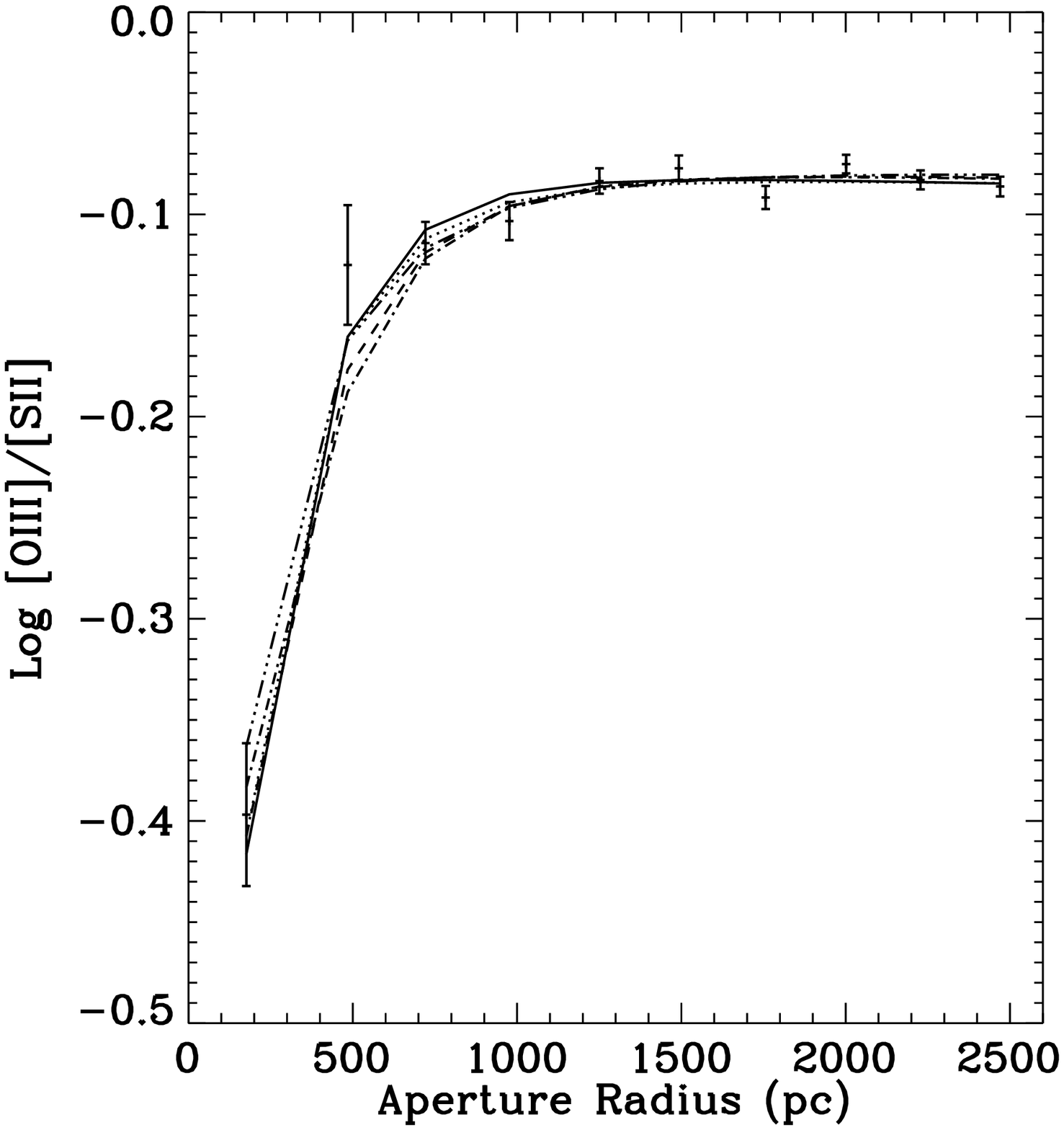}
\includegraphics[totalheight=0.35\textheight]{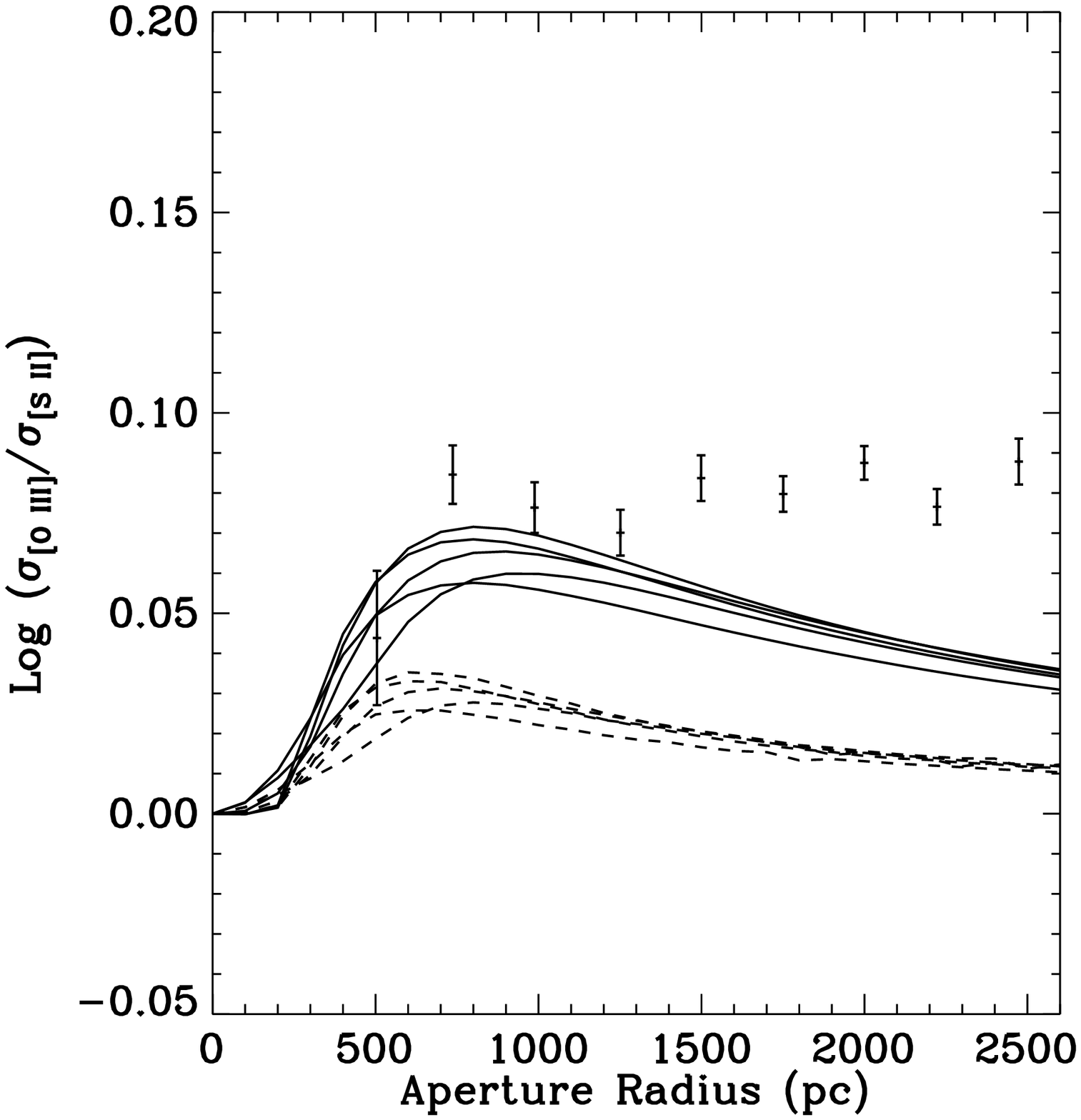}
\caption{Left panel: \oiii/\sii\ line flux ratio as a function of aperture radius for the data (points with error bars) and a few simple models. Right: The \oiii-to-\sii\ width ratio as a function of aperture radius for the data and the models. The two sets of curves indicate line widths measured by different methods, with solid lines indicate FWHM ratios and dashed lines indicate ratios from Gaussian fits. The models assume thin gas disk in circular rotation, with flat velocity dispersion.}
\label{fig:simuratio}
\end{center}
\end{figure*}

The left panel of Fig.~\ref{fig:simuratio} shows these models give reasonably good fits to the integrated line ratio profiles. The right panel of Fig.~\ref{fig:simuratio} shows the line width ratio as a function of aperture radius for these models.
 The solid curves and the dashed curves show the result for two different methods for measuring the line width. Because circular rotation makes more boxy profiles than Gaussian, the width measured from Gaussian fitting is different from that measured from FWHM. No matter how the width is measured, the width ratio between \oiii\ and \sii\ produced by all circular rotation models decreases strongly towards large apertures, inconsistent with what we observe. The decrease in width ratio towards large aperture is expected in the model since the line luminosity is increasingly dominated by flux from large radius and the line ratio profile flattens outward. Therefore, the line width ratio is probably not caused by the line ratio gradient and circular rotations. So, what is the real cause of the line width difference?

The line width is mainly contributed by two components: the thermal broadening and the bulk motion of the clouds. For gas at $T=10^4 {\rm K}$, the thermal broadening is approximately $15{\rm km/s}$ for \hal. For the red galaxies in our sample, the lines are very wide, with Gaussian sigmas ranging between 100km/s and 300km/s and a median of 176km/s in \nii\ lines. Therefore, the thermal broadening is a minor contributor. The width is likely dominated by bulk motion broadening or turbulence in the clouds.

To produce a width difference between different lines, we have to have
clouds with different line ratios and different line widths, and the
line ratio has to correlate with the line width. For example, for
\oiii\ and \sii\ to have different widths, we need a population of
clouds with high \oiii/\sii\ flux ratio and a population of clouds
with low \oiii/\sii. To make a wider \oiii\ than \sii , those high
\oiii/\sii\ clouds need to produce a wider width than those low
\oiii/\sii\ clouds.  Basically, to produce a constant width ratio with
radius, the width ratio has to be approximately the same everywhere in
the galaxy. Thus it requires at least two components of the ISM
throughout the galaxy that have different line ratios and different
line broadening. Because the total flux ratio has a gradient with
radius, the two components need to have approximately synchronous
line ratio gradient behaviours. One component is kinematically more
disturbed than the other and thus produces a wider line width. Our
observed width ratios indicate that the more disturbed component has
higher \oiii/\sii\ and \nii/\sii\ ratios, lower \sii/\hal\ ratio, and
similar \nii/\hal\ ratio to the more quiescent component.  Without
knowing the ionization mechanism for the gas, the origin of these
multiple components and the reason for their line ratio difference is
difficult to analyze.

\begin{table}
\begin{center}
\caption{Median line width ratios}
\begin{tabular}{c|c|c}
Line pair & Star-forming galaxies & Old red galaxies \\ \hline
$\sigma_{\nii}/\sigma_{\sii}$ &  $0.991\pm0.0004$ & $1.079\pm0.004$ \\
$\sigma_{\oiii}/\sigma_{\sii}$ & $1.006\pm0.001$ & $1.164\pm0.006$ \\ 
$\sigma_{\sii}/\sigma_{\hal}$ &  $1.033\pm0.0004$ & $0.933\pm0.004$ \\
$\sigma_{\nii}/\sigma_{\hal}$ &  $1.022\pm0.0003$ &$1.009\pm0.003$ 
\end{tabular}
\label{tab:medianratio}
\end{center}
\end{table}

\begin{figure}
\begin{center}
\includegraphics[totalheight=0.35\textheight]{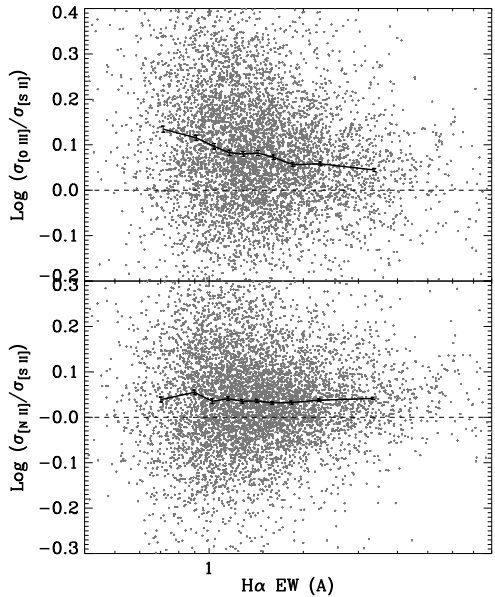}
\caption{Top panel: the width ratio distribution between \oiii\ and \sii\ as a function of \hal\ EW for line-emitting red galaxies. Bottom panel: same plot for the width ratio between \nii\ and \sii. }
\label{fig:lw_ew}
\end{center}
\end{figure}

We check if the line width difference changes with the line strength. Figure~\ref{fig:lw_ew} shows the width ratio distribution between emission lines as a function of \hal\ EW. We only use the brightest 25\% galaxies in total emission line luminosity and excluded among them those with uncertainty on line width ratio greater than 0.2 dex. The median \oiii-to-\sii\ and \nii-to-\sii\ width ratios always stay above 1. 
The \oiii-to-\sii\ width ratio seems to decrease slightly with increasing \hal\ EW. 

\begin{figure}
\begin{center}
\includegraphics[totalheight=0.35\textheight]{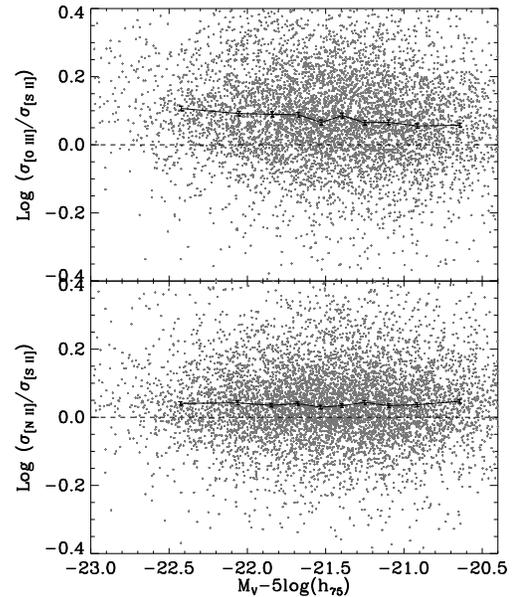}
\caption{Top panel: the width ratio distribution between \oiii\ and \sii\ as a function of luminosity for line-emitting red galaxies. Bottom panel: same plot for the width ratio between \nii\ and \sii. }
\label{fig:lw_lum}
\end{center}
\end{figure}

Figure ~\ref{fig:lw_lum} show the width radio distribution for the
line-emitting red galaxies as a function of absolute luminosity. The
same sample is used as in Fig.~\ref{fig:lw_ew}. The
\nii-to-\sii\ width ratio stays flat as a function of luminosity but
the \oiii-to-\sii\ width ratio declines slowly towards fainter
galaxies. The important point is that they always show significant
offset from zero in log space, suggesting that the reason causing the
width difference is universal in these galaxies.

\section{Discussion}\label{sec:discussion}

In this section, we first discuss what physical factors determine the emission line 
surface brightness profile, and show that the profile alone does not provide a 
discriminator between different ionization mechanisms. Then, we discuss the viability 
of the different ionization mechanisms in light of the observational results 
we presented above.

\subsection{Surface Brightness Profile} \label{sec:sbprofile}
Except shock heating and conductive heating by the hot gas, all other
major ionization mechanisms proposed involve photoionization. In this
section, we consider a generic photoionization model and examine which
of its parameters determine the emission line surface brightness
profile.

We assume that the ISM is filled with hot, ionized gas. Embedded in it
are neutral dense clouds. Each cloud is optically-thick to the
ionizing radiation. In photoionization equilibrium, for each cloud the
total emission line luminosity has to be equal to the total
photoionizing luminosity it receives, which is the incoming flux times
the projected area of the cloud. Therefore, the luminosity density
profile depends on the photoionizing flux profile (as a function of
radius) and the total projected cloud area per unit volume. For
example, for a volume filling factor of $f_g$, assuming the clouds
have an average volume of $\langle V \rangle$ and an average projected
area of $\langle A\rangle$, the luminosity density of line emission
would be
\begin{equation}
j(r) = F(r) {f_g(r) \over \langle V \rangle} \langle A \rangle  
\end{equation}
Here, $F(r)$ is the ionizing flux profile. The second term on the right hand 
side gives the number density of clouds. Multiplying 
it with the average projected area yields the total projected cloud area per unit volume.
To obtain the final surface
brightness profile, we also need to convolve the luminosity density
profile with the spatial distribution of the clouds. If the clouds all
reside in a disk with constant thickness, then the surface brightness
scales with radius in the same way as the luminosity density. However,
if the thickness of the disk increases with radius, like the Milky Way
gas disk, then the surface brightness profile would be much shallower.
Assuming the scale height of the disk is $H(r)$, the surface brightness of the 
line emission would be 
\begin{equation}
\Sigma (r) = F(r) f_g(r) H(r) {\langle A \rangle \over \langle V \rangle}
\end{equation}
The typical cloud area and volume could also change with distance from the center. Because the gas density ($n$) depends on distance from the galaxy center, if the mass distribution of the clouds is independent of the distance, then the typical $ \langle A \rangle/\langle V \rangle$  will scale as $n^{1/3}$. Therefore, the constraint from the surface brightness profile is 
\begin{equation}
\Sigma (r) = F(r) f_g(r) H(r) n^{1/3} \propto r^{-1.28}
\label{eqn:surfacebrightness}
\end{equation}

We do not have enough information about the geometry of the cloud
distribution and how the typical cloud sizes change with radius to
constrain the photoionizing flux profile.  Therefore, the extended
line emission only provides a partial constraint on the source of the
ionization. We need more information from other methods.

Many ionization mechanisms have been proposed to explain the observed emission 
line ratios in these red galaxies, which mostly have LINER-like line ratios. 
In the following sections we consider these mechanisms, dividing them
generically into three categories: a central photoionizing source,
distributed photoionizing sources, and shocks.

\subsection{Photoionization by an accreting SMBH}

An accreting supermassive black hole will emit X-rays and extreme UV
radiation that photoionizes surrounding gas clouds and produces line
emission. In this section we will examine the predictions of this
model for the line ratio gradients and line width differences.

\subsubsection{Line ratio gradients for a SMBH}  \label{sec:agn}

\begin{figure}
\begin{center}
\includegraphics[totalheight=0.19\textheight, viewport=-20 280 560 560, clip]{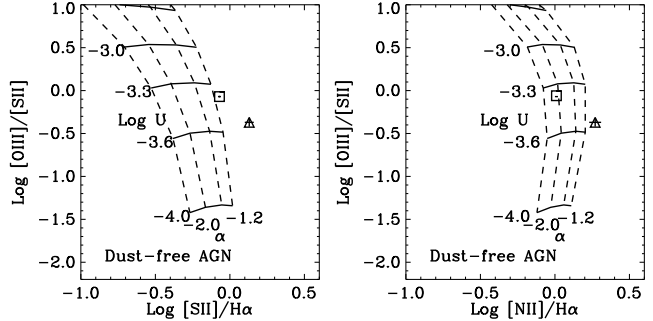}
\caption{Line ratio diagnostic diagrams for the center (triangle) and
  outskirts (square) of passive red galaxies, overlaid on the MAPPINGS III
  models of a classical dust-free AGN photoionization model. The
  models assume a metallicity of $2{\rm Z}_\odot$ and a hydrogen
  density of 1000 cm$^{-3}$. The solid lines indicate constant
  ionization parameters ($\log U = -4.0, -3.6,-3.3,-3.0$) and
  the dashed lines indicate constant spectral indices ($\alpha = -2.0,
  -1.7, -1.4, -1.2$).  }
\label{fig:mappings_agn}
\end{center}
\end{figure}

First, we will demonstrate that this model cannot explain the observed
line ratio gradients.  To do so, we use models calculated with the
MAPPINGS III codes \citep{DopitaS96,GrovesDS04I,Allen08}.
Figure~\ref{fig:mappings_agn} shows two line ratio diagnostic diagrams
with the grids representing the models presented by
\cite{GrovesDS04I}.

This standard photoionization model (with no dust or radiation
pressure) assumes a metallicity of $2{\rm Z}_\odot$ and a hydrogen
density of 1000 cm$^{-3}$. With the given range of parameters, this
model cannot produce the \sii/\hal\ and \nii/\hal\ ratios observed in
the center of these galaxies.  In fact, none of the dust-free
classical models in \cite{GrovesDS04I} can produce the central
\sii/\hal\ ratio: they are all too low.  It may be possible to fit the
\sii/\hal\ ratios by tweaking the N/S abundance ratio, or adopting the
dusty, radiation-pressure dominated photoionization model
\citep{Dopita02}. However, it is unlikely that the latter model is
applicable to the low intensity radiation fields in LINERs. 

Since we are not yet certain about the ionization mechanism, it is
premature to constrain the exact physical parameters using these
measurements. However, the models should provide a guide as to the
direction of change in the line ratios. Thus, we only use these models
to investigate what the line ratio gradients tell us about the change
in the physical parameters, not their precise values.

\begin{figure*}
\begin{center}
\includegraphics[totalheight=0.7\textheight, angle=90]{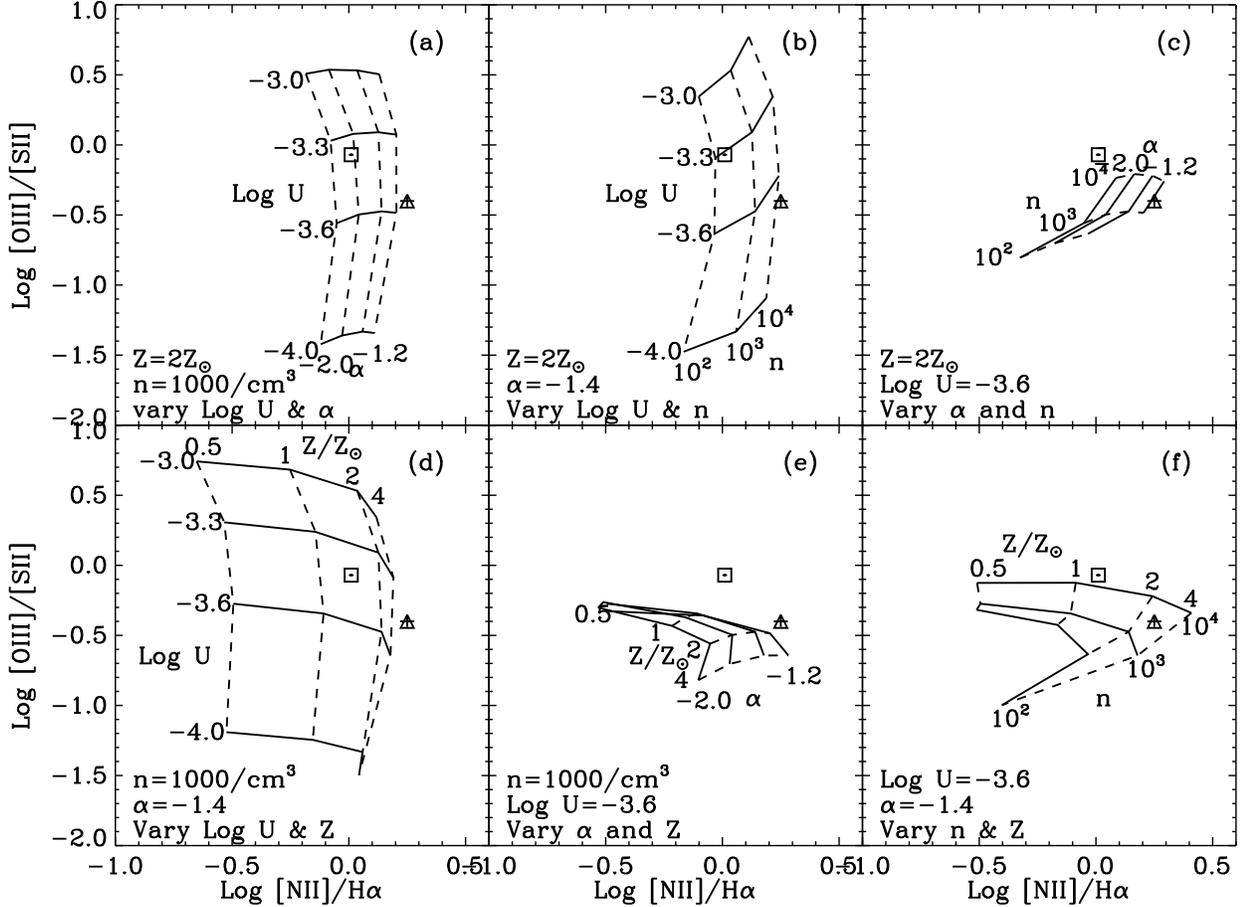}
\caption{\oiii/\sii\ vs. \nii/\hal\ for the classical dust-free AGN
  photoionization models (grids) and the measurements at the center
  (triangle) and outskirts (square) of line-emitting red galaxies. We
  explore the dependence of line ratios on four parameters: ionization
  parameter ($\log U$), spectral index ($\alpha$), gas density ($n$),
  and metallicity ($Z$). In each panel, we fix two parameters and vary
  the other two to see how the line ratios depend on each parameter.}
\label{fig:mapping_n2ha_all}
\end{center}
\end{figure*}

The line ratios are primarily determined by four parameters: gas
density ($n$), metallicity ($Z$), ionization parameter ($\log
U$)\footnote{$U$ is the dimensionless ratio of the ionizing photon
  flux density to the electron density, $U\equiv q(H^0)/(cn_H)$.}, and
the spectral index ($\alpha$). From the four strong lines, \sii, \nii,
\hal, \oiii, we have only three line ratios at each
position. Therefore, we cannot hope to determine the trend in all of
these parameters and have to start by keeping some parameters
fixed. In the six panels of Fig.~\ref{fig:mapping_n2ha_all}, we fix
two parameters at a time and look at the line ratio dependence on the
other two parameters, to determine all possible scenarios for the
observed line ratio variation. We only look at
\oiii/\sii\ vs. \nii/\hal\ diagram since the \sii/\hal\ data points
are not well covered by the models.

There are three ways that the \oiii/\sii\ ratio can increase outward:
an increase in the density (panel c), a decrease in the metallicity
(panel d), or an increase in the ionization parameter (panel a).  For
the first option, it is unphysical to expect the density to increase
outwards, which would require a higher gas pressure at larger radius
(since the temperature in the ionized gas is likely to be always near
$10^4{\rm K}$).

The second option, the metallicity gradient, is also not a likely
source for the change in \oiii/\sii\ ratio. At fixed density and
ionization parameter, it requires a factor of $\sim3$ change in
metallicity with radius, which is comparable though somewhat larger
than the observed stellar metallicity gradients
(\citealt{Kuntschner10}). However, the density is likely to decrease
outwards, which acts as a countervailing force and would require a larger metallicity
gradient. Indeed, since at low metallicity the \oiii/\sii\ ratio
becomes insensitive to $Z$, this possibility could be ruled out. 

The third option, an increasing ionization parameter, is much more
promising.  \oiii/\sii\ is quite sensitive to $\log U$: a variation of
more than 0.1 dex would dominate any other possible effect. Even
before considering specific models, we should suspect that the
ionization parameter in these objects increases outwards.

The ionization parameter is defined as the ratio between ionizing 
flux and gas density. Therefore, we now examine how gas
density changes with radius. X-ray observations of giant ellipticals have shown that the hot gas
density follows the square root of the stellar density profile, $n_e
\propto \rho_*^{1/2}$ (\citealt{MathewsB03} and references there
in). This means that the gas density profile falls with radius as
$r^{-p}$ with $0.5<p<1$ at the center and an increasing $p$ at large
radii. This range of central density slope is consistent with the
more recent X-ray measurements by \cite{Allen06}.
The temperature profile of the hot gas is much flatter, varying
by at most 50\% between the center and the outskirts
\citep{MathewsB03}. The gas clouds that generated the observed optical
line emission always have temperature near $10^4{\rm K}$. Therefore,
under pressure equilibrium, with the nearly constant temperature
profile in both the hot gas and the warm ionized gas, the density in
the warm ionized gas clouds should fall with radius in roughly the same way
as the hot gas density. 

It is important to note that this density scaling is only verified in
giant ellipticals. In fainter early-type galaxies it may not hold. 
Nonetheless, we expect the central density profile in faint ellipticals 
is also much shallower than $1/r^2$. Evidence for this comes from density 
profile measurements in the central regions ($\lesssim100{\rm pc}$) of a few 
fainter early-type galaxies (NGC 1052, NGC 3998, NGC 4579) from 
\sii\ line ratios. \cite{Walsh08} showed that the power-law indices
of their gas density profiles are around $-0.6$. Therefore, we use
the $n_e \propto \rho_*^{1/2}$ scaling as a working assumption here
and in the next section. Our main conclusion remains the same if one 
switches to a power law density profile with $n_e \propto r^{-1}$ or 
shallower.

In the case of a central ionizing source, the flux decreases as 
$r^{-2}$. Since the gas
density is at most decreasing as $r^{-1}$, the ionization parameter
must decrease outwards by a large amount, at least 0.5 dex. No change
in metallicity or spectral index could conceivably make up for this
decrease: this model inevitably predicts a strongly decreasing
\oiii/\sii\ ratio with radius, the opposite of what we observe.
Therefore, unless the gas density profile actually falls faster than
$r^{-2}$, the AGN photoionization model cannot explain the observed
line ratio gradients.

Meanwhile, there are also three ways for the \nii/\hal\ ratio to
decrease outward: a softening of the ionizing spectrum (panel a), a
decrease in the density (panel b), a decrease in the metallicity
(panel d), or a combination of these. The outward decreasing density
provides a natural solution. Though its decline with radius might be
too slow to explain all the change in \nii/\hal\ ratio.
An additional
contribution from metallicity gradient and a change in the spectral
index might be needed as well. With the current modeling uncertainty,
we cannot break the degeneracy among these possibilities.

\subsubsection{Comparison to line width differences observed in nearby Seyferts} 

Next, we consider the observed variations of line width between our
various lines. We conclude here that the variations we observe are
probably not related to those known to exist for Seyfert galaxies.

Velocity width variations among different emission lines have been
observed in classical nearby Seyferts and LINER nuclei
\citep{FilippenkoH84,Filippenko85,deRobertisO86,HoFS96}. In most
cases, the line widths correlate strongly with critical density for
collisonal deexcitation: lines with higher critical density tend to
have larger widths. In a minority of Seyfert 2s, the line width
correlates with the ionization potential of the ions.  

The line width differences we observe for LINERs is broadly consistent
with that seen in local Seyferts. In a sample of 18 Seyfert 2
galaxies presented by \cite{deRobertisO86}, the median of \nii-to-\sii\ width ratio is $1.11\pm0.06$, the median of \oiii-to-\sii\ width
ratio is $1.18\pm0.08$. Our width ratios are only slightly smaller.

\cite{FilippenkoH84} proposed the following picture to explain the
line width differences. The ionization flux decreases outward
according to the inverse-square law. If the density also falls as
$r^{-2}$, then the ionization parameter seen by each cloud will be the
same. If all clouds are optically-thick to the ionizing radiation,
then they will all have the same ionization structure and produce the
same set of emission lines. The relative line ratio will vary
according to the density of each cloud. Lines with a high critical
density will be mainly contributed by a high density clouds, which are
closer to the nucleus and have higher velocities.

In this scenario, the \nii/\sii\ and \oiii/\sii\ flux ratios should
increase towards the center, the opposite of what we observe for
LINERs. In addition, for most red galaxies the line emission is
spatially extended, with an average surface brightness profile falling
as $r^{-1.28}$. Thus, the line luminosity is not dominated by the
central regions where the density gradient is steep.  The kinematic
structure on large scales is also different from the Keplerian
rotation found near the SMBH. Therefore, while the scenario is
applicable to the narrow line regions of Seyferts, cannot be
applicable in our case. The similar line width ratios of our results
and those of \cite{deRobertisO86} may simply be a coincidence.

%\begin{table}
%\begin{center}
%\caption{Critical densities and ionization potentials}
%\begin{tabular}{c|c|c}
%Lines & Critical density (${\rm cm}^{-3}$) & Ionization potential (eV) \\ \hline
%\oiiiw\ & $6.8\times10^5$ & 35.1  \\
%\niiw\ & $6.6\times10^4$ & 14.5 \\
%\siiaw\ & $1.4\times10^3$ & 10.4 \\
%\siibw\ & $3.6\times10^3$ & 10.4
%\end{tabular}
%\label{tab:lineprop}
%\end{center}
%\end{table}

\subsection{Photo-ionization by Distributed Ionizing Sources}

The suggestion that the ionization parameter increases outwards
naturally points to a slower decrease in the flux and thus to
distributed ionizing sources rather than a central one.  A number of
models have been proposed along these lines, such as photoionization
by hot evolved stars and by the hot X-ray emitting gas. In this
section we first discuss the generic predictions of models with
distributed sources, and then discuss particular models in more
detail.

\subsubsection{Line ratio gradients from distributed sources}

Although the ionizing spectra produced by these models differ from
that produced by an AGN, the overall dependence of line ratios on
ionization parameter and metallicity are very similar. Thus, as in the
case of AGN, to explain the line ratio gradients we need the
ionization parameter to increase outwards. Distributed sources can
produce this trend.

Suppose the ionizing sources are distributed like the stars, i.e.,
their luminosity density profile follows the stellar density
profile. Then, we can compute the ionizing flux profile using the
latter. Assuming that the galaxy is spherically symmetric, that the
stellar density profile is $\rho(r)$, and that the average number of
output photoionizing photons per unit time per unit stellar mass is
$Q_0$, the total integrated ionizing flux at distance $D$ from the
center of the galaxy is:

\begin{align}
F(D) &= \int_0^{\infty} \mathrm{d}r \int_0^{2\pi} \mathrm{d}\phi 
        \int_0^{\pi} {Q_0 \rho(r) r^2 \sin\theta \over 4\pi (D^2+r^2 - 2Dr \cos\theta)} \mathrm{d}\theta  \\
     &= {Q_0 \over 2} \int_0^{\infty} \rho(r) {r\over D} \ln { D+r \over |D-r|} \mathrm{d}r  \label{eqn:fluxint}
\end{align}
The integral can be solved by substituting $r=D(1+e^u)$ for the $r>D$ part and substituting $r=D(1-e^u)$ for the $r<D$ part.

For the stellar density profile, we adopt the $\gamma$-model described by \cite{Dehnen93}:
\begin{equation}
\rho(r) = {(3-\gamma) M \over 4\pi} {a \over r^\gamma (r+a)^{4-\gamma} } 
\label{eqn:dehnen}
\end{equation}
where $M$ is the total mass, and $a$ is a scaling factor which relates to 
the effective radius $R_e$, depending on the inner slope, $\gamma$. The
$\gamma=1$ model correponds to the \cite{Hernquist90} profile, the
$\gamma=2$ model corresponds to the \cite{Jaffe83} profile, and the
$\gamma=1.5$ yields the best approximation of the de Vaucouleurs'
$R^{1/4}$ model in surface brightness profile. Putting this model in
Eqn.~\ref{eqn:fluxint}, we integrate numerically to obtain the total
ionizing flux as a function of radius for a model galaxy with an
effective radius of 4.8 kpc and a stellar mass of
$7.5\times10^{10}M_\odot$, the medians for the top 25\% line-emitting
passive red galaxies in our sample.

\begin{figure}
\begin{center}
\includegraphics[totalheight=0.35\textheight]{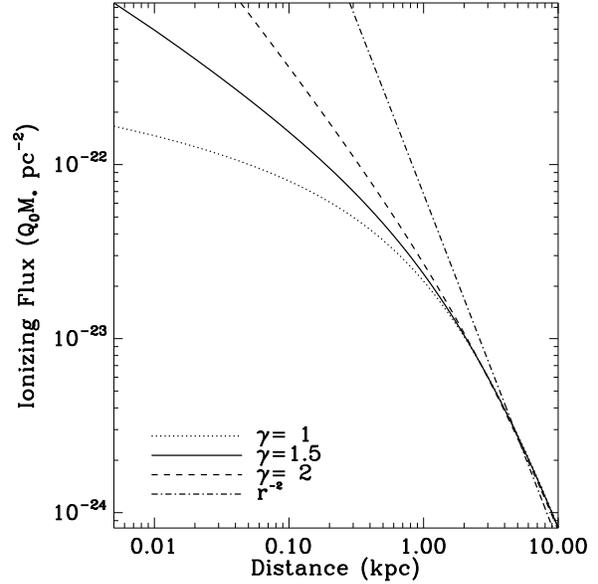}
\caption{The integrated ionizing flux from a system of ionizing sources following the stellar density profile as a function of distance to the center of a model galaxy. The curves correspond to different $\gamma$-models for the stellar density profile. The solid line corresponds to the model with the best fit to the de Vaucoulers' profile in surface brightness. The long dashed line corresponds to the inverse square law as expected in the AGN model. All models show flatter flux profile.} 
\label{fig:pagbflux}
\end{center}
\end{figure}

Figure~\ref{fig:pagbflux} shows the resulting ionizing flux profile for three models with different $\gamma$ values, along with the inverse square law expected from a central ionizing source. In the inner kpc, distributed ionizing sources will produce a much shallower ionizing profile than the inverse square law. 

\begin{figure}
\begin{center}
\includegraphics[totalheight=0.35\textheight]{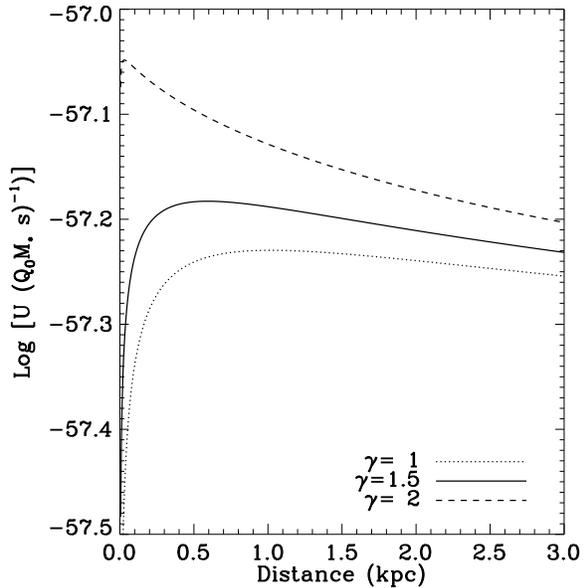}
\caption{The ionization parameter produced by a system of ionizing sources following the stellar density profile in a galaxy shining a cloud as a function of distance from the galaxy center. We assumed gas density profiles of $n(r) \propto n_*^{1/2}$ and normalize them to $100 {\rm cm}^{-3}$ at 1 kpc. The three curves correspond to three different $\gamma$-models as described by Eqn.~\ref{eqn:dehnen}. $M_*$ is the stellar mass of the galaxy.
} 
\label{fig:pagblogu}
\end{center}
\end{figure}

In Figure~\ref{fig:pagblogu}, we divide the ionizing flux profiles by
a gas number density profile to see how the dimensionless ionizing
parameter will vary with radius under these different models. We adopt
a gas density profile that scales as the square root of the stellar
density profile, $n_g \propto n_*^{1/2}$ (\citealt{MathewsB03} and
references therein, also see discussion in \S\ref{sec:agn}), and normalize them to $100{\rm cm}^{-3}$ at 1
kpc. This warm gas density is consistent with observations \citep{HeckmanBvB89, DonahueV97} 
and our assumption of pressure equilibrium between the warm gas ($T\sim10^4K$) 
and the hot gas ($T=10^6$ -- $10^7{\rm K}$, $n=0.1$ --$1 {\rm cm^{-3}}$).
Interestingly, for the $\gamma=1.5$ model, which gives the best
fit to de Vacucouleurs' profile, the ionizing parameter displays the
same trend as we observe, as shown by the \oiii/\sii\ ratio profile in
Fig.~\ref{fig:lineratio_ring_data}. It not only produces the increase
with radius in the central part but also a slow decline on the
outskirts.

\subsubsection{Luminosity dependence}

A prediction of the above model is that the line ratio gradient should
have a luminosity dependence. Many studies \citep{Lauer95, Faber97,
  Rest01, Ravindranath01, Lauer05, Ferrarese06, Glass11} have shown
that the inner power-law slope of the stellar luminosity density
profiles changes from $-1$ for bright galaxies to $-2$ for faint
galaxies. As shown by Figure~\ref{fig:pagblogu}, these different
stellar density profiles should generate different gradients in the
ionization parameter. The transition point is approximately at $M_B =
-20.5$ (or around $M_V = -21.3$). Here, we investigate whether the
gradients depend on luminosity in the expected manner.

We divide our passive red galaxy sample at $M_V=-21.3$ into bright 
and faint samples to look for the luminosity dependence.
The bright sample has a median stellar
mass of $1.0\times10^{11}M_\odot$ and a median effective radius of
$5.8~{\rm kpc}$. For the faint sample, the corresponding numbers are
$4.7\times10^{10}M_\odot$ and $3.4~{\rm kpc}$.

\begin{figure}
\begin{center}
\includegraphics[totalheight=0.35\textheight]{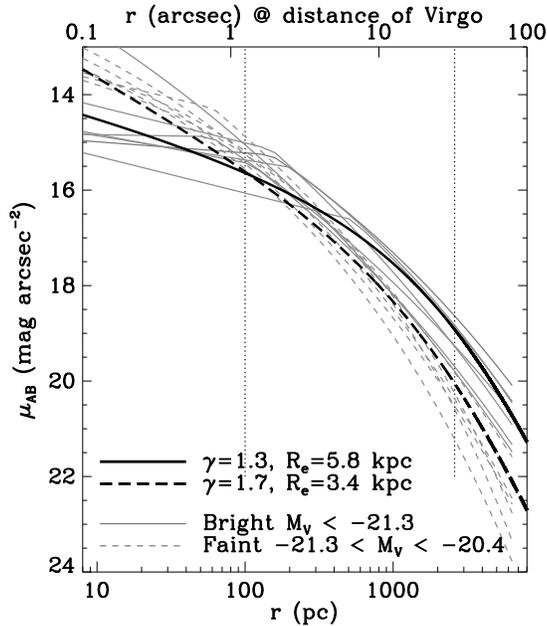}
\caption{Surface brightness profiles for the 14 early-type galaxies in the Virgo 
cluster that would satisify our luminosity cut ($M_V<-20.4$), as fit by \cite{Ferrarese06}.
The solid gray curves and the dashed gray curves represent galaxies brighter and fainter 
than $M_V$ of $-21.3$, respectively. The two thick curves represent the profiles
computed for two \cite{Dehnen93} models, with parameters indicated in the legend. 
They provide reasonable approximations to the profile difference between bright and
faint early-type galaxies. The two vertical dotted lines indicate the range of apertures
probed in this paper.}
\label{fig:virgogals}
\end{center}
\end{figure}

First, we show that \cite{Dehnen93} models with different $\gamma$ values
can provide reasonable approximations to the density profiles 
of bright and faint galaxies in our sample.
To demonstrate this, we compare the models to the surface brightness
profile fits presented by \cite{Ferrarese06} for the 14 early-type 
galaxies in the Virgo cluster that satisfy our luminosity cut ($M_V < -20.4$).
In Fig.~\ref{fig:virgogals}, the gray solid curves show profiles of the brighter galaxies 
with $M_V<-21.3$ and the gray dashed curves show those of the fainter ones.
The fainter galaxies generally have steeper profiles, despite their smaller
Sersic indices as reported by \cite{Ferrarese06}. The difference
can be reasonably approximated by the difference between the 
surface brightness profiles of two \cite{Dehnen93} models with $\gamma=1.3$, $R_e=5.8~{\rm kpc}$,
and with $\gamma=1.7$, $R_e=3.4~{\rm kpc}$.  

\begin{figure}
\begin{center}
\includegraphics[totalheight=0.35\textheight]{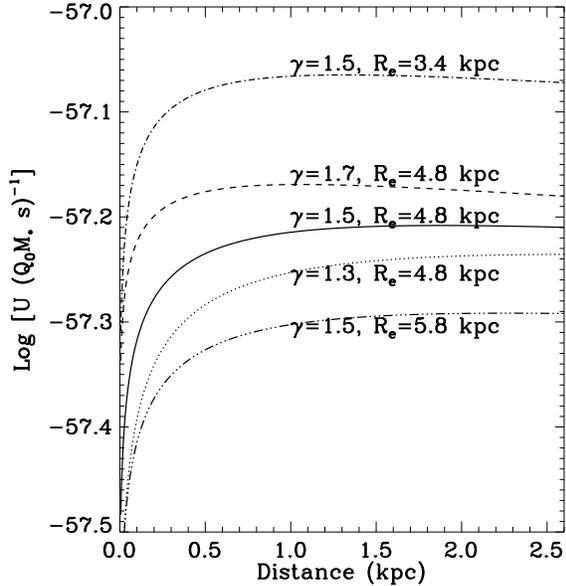}
\caption{Luminosity-weighted ionization parameter within aperture as a function of aperture radius, for five models with different stellar density profiles. The density profiles are specified by its inner power-law slope, $\gamma$, and effective radius, $R_e$. The gas density are assumed to scale as the square root of the stellar density and is normalized to $100 {\rm cm}^{-3}$at 1 kpc for all models. This figure shows that $\gamma$ controls the shape of the resulting ionization parameter profile while $R_e$ affects mainly the normalization. 
}
\label{fig:logu_lumdep}
\end{center}
\end{figure}

Next, we demonstrate that the shape of integrated ionization parameter profile 
is not sensitive to the effective radius of the galaxy, but is primarily controlled by 
the $\gamma$ parameter.
Bright galaxies not only have shallower inner density profiles, but also have 
larger effective radii than faint
galaxies. In Figure~\ref{fig:logu_lumdep}, we plot models with various
choices for $\gamma$ and $R_e$.  The ionization parameter plotted is
the luminosity-weighted average within an aperture. For the luminosity
weighting, we assume a power-law surface brightness profile with an
index of $-1.23$, as derived from a fit to the \oiii\ profile in
Fig.~\ref{fig:linelum12_z}\footnote{The data show that the line
  emission surface brightness profile has only a weak dependence on
  the galaxy luminosity. Thus, we adopt the same luminosity profile
  for all models.}. Combined with the ionization parameter profile
from the model, we compute the luminosity-weighted average ionization
parameter as a function of aperture radius. In all models, the gas
density is assumed to scale as the square root of the stellar density
and they are all normalized to be $100 {\rm cm}^{-3}$ at 1 kpc.

The middle three curves in Fig.~\ref{fig:logu_lumdep} are models with
the same $R_e=4.8$ kpc but different $\gamma$. They have different
slopes in both the outer part and the inner part. They differ little
in overall normalization. The top and bottom curves have the same
$\gamma$ as the solid curve, but differ in $R_e$. These three cases
have nearly identical shapes, but differ significantly in their
normalization. Other factors can impact the normalization, including
$Q_0$, the stellar mass, and the normalization of the gas
density. Without knowing $Q_0$ and the normalization of the gas
density, the normalization of the curve is free to vary.  In contrast,
only $\gamma$ controls the shape of the ionization parameter
profile. Since we know $\gamma$ varies with luminosity, a definite
prediction by this model is that the bright sample and faint sample
should differ in the shape of their integrated \oiii/\sii\ profile.

\begin{figure}
\begin{center}
\includegraphics[totalheight=0.35\textheight]{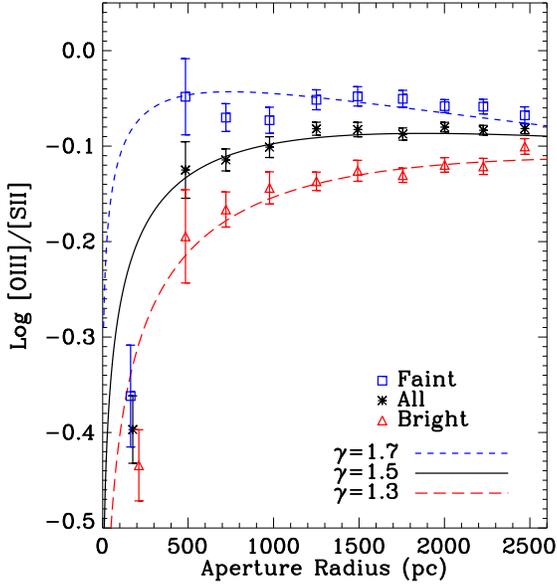}
\caption{The aperture \oiii/\sii\ ratio as a function of aperture size for the whole sample (stars), bright sample (triangles, $M_V < -21.3$), and faint sample (squares, $M_V > -21.3$). The curves represent predictions of three simple models with different $\gamma$ parameter (Eqn.~\ref{eqn:dehnen}). They are {\it not} fits to the data, but a scaled and shifted version of the luminosity-weighted average ionization parameter. See text for detail.
}
\label{fig:o3s2_logu_all}
\end{center}
\end{figure}

{Now we check this luminosity dependence in the data. We divide 
all passive red galaxies into bright and faint subsamples at $M_V=-21.3$.}
As in the whole sample, we
select the top 25\% galaxies in each subsample that have the brightest
total emission line luminosity. 
Figure~\ref{fig:o3s2_logu_all} shows the \oiii/\sii\ ratios as a
function of aperture size for the bright sample and the faint sample
separately, along with those for the whole sample. The curves
represent the prediction of three models with different $\gamma$
parameter, $R_e$ and stellar mass, with the latter two parameters
adopting median values in the data. The models are calculated in the
same way as for Fig.~\ref{fig:logu_lumdep}. 

To convert the luminosity-weighted average ionization parameter, we
used the median stellar mass and $R_e$ for each model,
assumed that $Q_0$ was constant, and assumed that every 0.3 dex in
$\log U$ translates to 0.5 dex in \oiii/\sii\ ratios.  Then we shifted
the models vertically by varying the gas density normalization so that
the $\gamma=1.3$, 1.5, and 1.7 models match roughly the normalization
of the data points for the bright sample, the whole sample, and the
faint sample, respectively.  We did not perform an explicit fit using
these data because there are still too many poorly-known factors in
the model.  Under our assumptions, the gas density at 1 kpc must be
$\sim13\%$ lower for the faint sample than in the full sample, and
$\sim6\%$ higher for the bright sample. This normalization difference
between the bright and faint samples does not have to be due to a density 
difference. It could also be due to the higher fraction of flat systems 
(lenticular galaxies) in the faint sample \citep{Bernardi10}. The stars in an intrinsically
flat galaxy would be systematically closer to the gas and yield a larger
ionization parameter, hence higher \oiii/\sii\ ratios.

Although the normalizations match the data by design, the shape of the
models are set completely by the $\gamma$ values. The bright galaxy
sample has an increasing \oiii/\sii\ ratio (integrated) with radius,
matching the model prediction of a stellar density profile with a
flatter inner slope (small $\gamma$). The faint galaxies have a much
flatter outer slope, matching the prediction by a stellar profile with
a steeper inner slope (large $\gamma$). In addition, the fainter
sample displays a much steeper line ratio gradient on small radii
($<500~{\rm pc}$) than the brighter sample, matching the general trend
predicted by the model. 

The data on the smallest scales do not match the data. This mismatch
could be due to the overly simplified assumptions we made. The scaling
between \oiii/\sii\ ratio and $\log U$ may be non-linear; the inner
gas density profile may be steeper than assumed. These could all
change the shape of the curves. 

It is remarkable that a simple single-parameter model is able to
predict the overall luminosity dependence of the \oiii/\sii\ profile
shape in the data.  It provides a strong support for ionization
mechanisms invoking sources that are distributed like the stars.

\begin{figure}
\begin{center}
\includegraphics[totalheight=0.35\textheight]{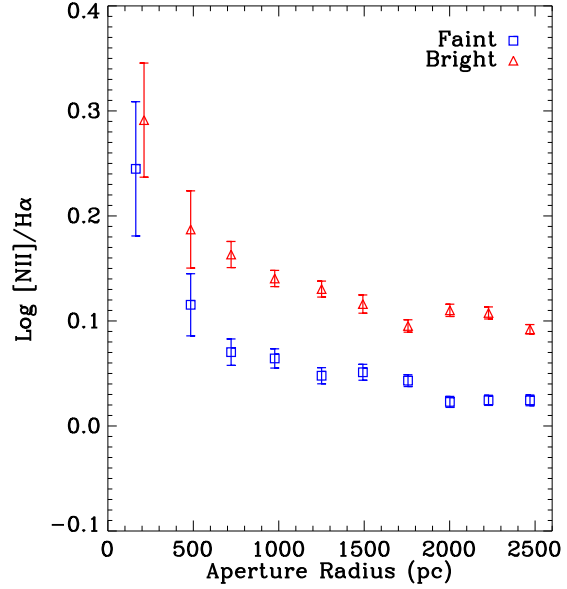}
\caption{Aperture \nii/\hal\ ratios for the bright (triangles) and faint (squares) passive red galaxies as a function of aperture radius. 
}
\label{fig:n2ha_logu_all}
\end{center}
\end{figure}

In Fig.~\ref{fig:n2ha_logu_all}, we show the \nii/\hal\ ratios as a
function of aperture size for the bright and faint samples. The
brighter sample always shows larger median \nii/\hal\ ratios than the
fainter sample. This correlation between \nii/\hal\ and galaxy luminosity 
has been seen by \cite{Phillips86}. Our result shows that this correlation 
exists at all aperture scales. As we learned from 
Fig.~\ref{fig:mapping_n2ha_all}, to
increase \nii/\hal\ with photoionization, we have to either increase
the metallicity, increase the density, or use harder ionizing
spectra. The density is unlikely to vary by more than a factor of 10
between the bright galaxies and faint galaxies. And if all galaxies
are powered by the same ionizing sources, the spectra should also be
the same. Therefore, the difference in \nii/\hal\ is most likely due
to the gas-phase metallicity difference between the two samples.

To summarize, photoionization by distributed ionizing sources
following the stellar density profile can naturally produce the
general variation of ionization parameter with radius, including the
sharp rise at small radius and gentle decline on large radius. It is
also able to produce the overall direction of the
luminosity-dependence of the line ratio gradient. This strongly
indicates that the spatial distribution of the true ionizing source is
similar to the stellar distribution.

\subsubsection{Post-AGB stars}

\cite{Binette94} proposed that photo-ionization by post-AGB stars
could explain the extended line emission in red galaxies.  The spatial
distribution of these post-AGB stars should be very similar to the
overall stellar distribution. Therefore, based on the results of the
previous sections, the diffuse ionizing field they form can produce
the observed line ratio gradient and its luminosity dependence. In
this section we discuss relevant aspects of post-AGB evolution and
planetary nebulae, and evaluate whether they can produce enough
ionizing photons and a sufficiently high ionization parameter.

These stars have left the asymptotic giant branch and are evolving
horizontally on the H-R diagram towards very high temperatures
($\sim10^5$K) before cooling down to form white dwarfs. They are
burning hydrogen or helium in a shell around a degenerate
core. Because their temperatures are high enough to ionize the
surrounding medium and plenty of material has been expelled from them
in earlier evolutionary stages, they are often accompanied by a
planetary nebula. After their planetary nebulae disperse into the
interstellar medium, the long-lived post-AGB stars can produce a
diffuse ionizing field.

Most of our knowledge about post-AGB stars comes from studies of
planetary nebulae. Observations of planetary nebulae have shown their
dynamical ages are about 30000 years \citep{Schonberner83,
  Phillips89}. The time spent by stars in the post-AGB phase is a very
strong function of their core mass \citep{Renzini83}. For high mass
post-AGB stars, the evolution is very fast: a post-AGB star with core
mass of $1.0M_\sun$ will have a nuclear burning time of only 25 yr
\citep{Tylenda89} and fades by a factor of 10\ in luminosity on a
similar timescale. These stars evolve too fast to make their planetary
nebulae visible for long. Thus most planetary nebulae have central
stars with core masses less than $0.64M_\odot$ \citep{TylendaS89}. On
the other hand, very low mass post-AGB stars ($M < 0.55M_\sun$) evolve
so slowly that before they raise their temperature to $3\times10^4K$
(needed to ionize hydrogen) the material expelled in the AGB phase has
completely dissipated into the interstellar medium. These stars,
termed `lazy post-AGB' \citep{Renzini81}, will not appear as planetary
nebulae. Therefore, considering the lifetime of post-AGB stars and the
dynamical ages of the planetary nebulae, the central stars of
planetary nebulae have to be post-AGB stars with core mass in a narrow
range of $0.55$---$0.64M_\sun$ (\citealt{TylendaS89}, also see
\citealt{Buzzoni06} Fig. 15 for a nice illustration of the mass
dependent PN visibility). This core mass range corresponds roughly to
an initial mass of $1$---$3M_\sun$ \citep{Weidemann00}.  Lazy post-AGB
stars and those post-AGB stars that live longer than 30000 years will
form a diffuse ionizing field capable of ionizing the neutral gas in
the larger scale interstellar medium.

However, our current understanding of the late stages of stellar
evolution is fairly poor and we do not know the temperature and age
distribution of these stars well. Most post-AGB stars observed are
either hidden inside planetary nebulae or observed when they are not
yet hot enough to ionize the nebula. Few hot naked post-AGB stars have
been observationally identified \citep{Napiwotzki98, Brown00,
  Weston10}, which may be due to strong observational bias since they
will be very luminous in the extreme UV but very faint in the
optical. Therefore, it is uncertain what fraction of post-AGB stars
contribute to the large-scale photo-ionizng field and what fraction
are hidden inside planetary nebulae.

Fortunately, there are two clues indicating that planetary nebulae do
not dominate the line luminosity in most of our line-emitting
galaxies. First, if the line emission is dominated by planetary
nebulae, their kinematics should follow the stellar kinematics
exactly. However, \cite{Sarzi06} showed that the ionized-gas
kinematics is decoupled from the stellar kinematics in the majority of
galaxies in their sample. 

Second, we can estimate the total \oiii\ luminosity contributed by
planetary nebulae by integrating the planetary nebula luminosity
function. We take the double exponential function given by
\cite{Ciardullo89},
\begin{equation}
\log N(M) = 0.133M + \log [1 - e^{3(M^*-M)}] + const,
\end{equation}
where M is defined as $M_{\rm [OIII]} = -2.5\log F_{\rm
  [OIII]}-13.74$. The bright cut-off magnitude $M^*$ is -4.47. The
faint cut-off magnitude is 8 mags fainter than $M^*$
\citep{Henize63}. The normalization is usually given as the total
number of planetary nebulae within the two cut-off manitudes divided
by the total luminosity of the galaxy. We adopt the median value
reported by \cite{Buzzoni06} for a sample of early-type galaxies,
which is $N=1.65\times10^{-7}L_{gal}/L_\odot$. With these, we found
the total \oiii\ luminosity produced by planetary nebulae should be
$L(\oiii)= 1.35\times10^{28} L_{\rm gal}/L_\odot~{\rm erg~s}^{-1}$. To
estimate the total PN light we observed through the fiber, we should
use the fiber magnitude to derive $L_{\rm gal}$. For the 25\% passive
red galaxies at $0.09<z<0.1$ with the brightest total line luminosity,
the median V-band absolute magnitude within the fiber aperture
(derived using the fiber mags) is $-20.09$. This produces a median
\oiii\ luminosity of $1.21\times10^{38} {\rm erg~s^{-1}}$, much
smaller than the median \oiii\ luminosity among these galaxies, which
is $6.2\times10^{39} {\rm erg~s^{-1}}$. Therefore, we conclude that
planetary nebulae is a minor contributor to the total line emission in
these galaxies.

Can the diffuse ionizing field produced by naked post-AGB stars explain what we see? In this case, the sources are distributed like the stars and will produce the observed ionization parameter gradient and its luminosity dependence. The question is whether there are enough ionizing photons and what ionization parameter they can produce. 

\cite{Binette94} argue that the diffuse ionizing field has many more
ionizing photons than planetary nebulae can produce, especially when
the stellar population gets older than 3 Gyr. They estimate a total
$Q_0\sim1\times10^{41} s^{-1} M_\odot^{-1}$. The median stellar mass
with the fiber aperture for the 25\% passive galaxies at $0.09<z<0.1$
with the brightest total line luminosity is $2.1\times10^{10}
M_\odot$. Assuming all post-AGB ionizing photons are completely
absorbed and on average it takes 2.2 photoionizing photons to produce
one \hal\ photon, this will produce a median \hal\ luminosity of
$2.9\times10^{39} {\rm erg s}^{-1}$, which is about 1/3 of the median
observed \hal\ luminosity $8.3\times10^{39} {\rm erg s}^{-1}$ (before
extinction correction, but we expect the extinction to be
small). Considering the uncertainties on $Q_0$ and other parameters
involved in the calculation, this can be considered as a good
agreement. This luminosity is much larger than the contribution from
all planetary nebulae.  More detailed calculations by
\cite{Stasinska08} and \cite{CidFernandes11} yield similar
results. Thus, post-AGB stars can produce enough ionizing photons to
account for most of the line luminosity, as long as all these photons
are trapped inside the galaxy. This latter question is related with
how the gas clouds are distributed relative to the post-AGB stars.

In light of the decoupling of the gas kinematics from the stellar
kinematics, let us assume that the line-emitting gas clouds are
randomly distributed in relation to the post-AGB stars. In this case,
we can estimate the ionization parameter by multiplying the $Q_0$ for
post-AGB into Fig.~\ref{fig:pagblogu}. This yields an ionization
parameter of $\log U= -5.2$ at 1 kpc, a factor of 10 lower than what
is required ($\log U \sim -3.5$ from Figure
\ref{fig:mapping_n2ha_all}, or -4 according to \citealt{Binette94}).
Therefore, although there may be enough photons from post-AGB stars to produce
the total line luminosity, in our model the light is deposited onto
clouds far away from the ionizing sources. In consequence, the flux is
significantly reduced and the resulting ionization structure and line
ratios are very different from the expectation from clouds closer to
the ionizing sources.

A possible solution might be that clouds closest to individual 
post-AGB stars dominate in luminosity; if this were so, because they 
also have the highest ionization parameters, the luminosity-weighted 
average ionization parameter among all clouds would be raised. Here 
we determine that this solution is unlikely to work. 
To evaluate this possibility, we compute the average spacing between
post-AGB stars using a rough number density computed by dividing the
total ionizing flux by that from an average star. We use an individual
post-AGB luminosity of $10^4L_\odot$, significantly higher than
average, which maximizes the contribution of individual stars in this
calculation. With this luminosity, the inter-spacing is around $85~{\rm
  pc}$ at $r=1~{\rm kpc}$ and increases outwards.
For gas clouds that are randomly distributed with regard to the
post-AGB stars, 
the luminosity each cloud receives is Flux$\times$Area. The Flux
consists of two components, the diffuse background $F_b$, and the flux
from its nearest post-AGB star $Q_1/(4\pi r^2)$. Over a spherical
volume with diameter equal to the inter-post-AGB spacing ($r_{\rm
  max}$), the total luminosity due to the diffuse background is
\begin{equation}
L_{bkgd} = F_b {4\pi \over 3} r_{\rm max}^3 n_c \langle A\rangle,
\end{equation}
where $n_c$ is the number density of gas clouds and $\langle A\rangle$ is the average projected cloud area. The total luminosity due to an individual post-AGB star is 
\begin{align}
L_1 &= (\int_0^{r_{\rm max}} {Q_1 \over 4\pi r^2} 4\pi r^2 \mathrm{d}r ) n_c \langle A\rangle \\
    &= Q_1 r_{\rm max} n_c \langle A\rangle
\end{align}
where $Q_1$ is the total photoionizing photon output rate of the star.

\begin{figure}
\begin{center}
\includegraphics[totalheight=0.35\textheight]{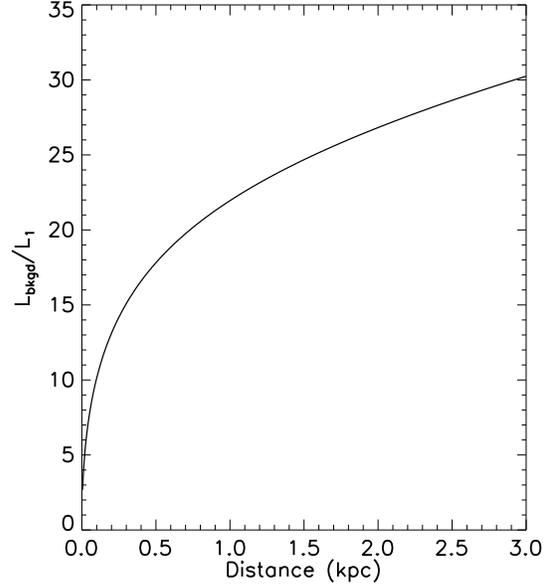}
\caption{The ratio between emission line luminosity produced by
  diffuse pAGB ionizing background shining on randomly distributed
  clouds and that produced by a single nearby pAGB star as a function
  of radius. This indicates that the diffuse background is dominant in
  ionizing randomly distributed clouds, confirming the predicted
  ionization parameter.}
\label{fig:pagbbkgd}
\end{center}
\end{figure}

Figure~\ref{fig:pagbbkgd} shows the ratio $L_{bkgd}/L_1$ as a function
of radius for the $\gamma=1.5$ model. Except for the very central part
of the galaxy, the luminosity due to the background is significantly
larger than that due to individual nearby post-AGB stars. This result
suggests the ionizing field produced by post-AGB stars is fairly
smooth in most parts of the galaxies. The luminosity-weighted
ionization parameter should be fairly close to what we showed in
Fig.~\ref{fig:pagblogu}. Granularity in the ionization field is thus
unlikely to cause a substantially increased ionization parameter.

Increasing the ionization parameter requires the clouds to be closer
to the post-AGB stars. A factor of 4 decrease in average distance
would probably be enough to bring the ionization parameter into the
right ballpark. To achieve this, either the clouds must originate from
progenitors of the post-AGB stars or the post-AGB stars must be
preferentially distributed near the warm/cool gas. Since the gas is
quite often kinematically decoupled from the stars, both scenarios
require that the post-AGBs share the same origin as the gas, rather
than that of the main stellar population. Among all post-AGB stars,
those with the largest core mass dominate in luminosity, which are
also the youngest. If both the gas and the dominant post-AGB stars are
associated with the most recent star formation episode, the dominant
post-AGB population might share a similar spatial distribution and
kinematics with the gas. This scenario would help resolve the deficit
in ionization parameter. However, because our sample selection
involves a cut in $D_n(4000)$ which would exclude systems with 
more than a few percent\footnote{The exact mass fraction of the young population
tolerable by our $D_n(4000)$ cut depends on the assumptions used in
the stellar population modeling. Assuming an old simple stellar population
with solar metallicity and an age of 4.6 Gyr, which yields the observed
median $D_n(4000)$ of 1.9, our $D_n(4000)$ cut at $z\sim0.1$ can only
tolerate at most $2\%$ of its stellar mass coming from a population younger
than 1 Gyr.} of their stellar mass from a young stellar 
population ($<1 {\rm Gyr}$), we consider this scenario unlikely.
Nonetheless, it can be tested by looking at planetary
nebulae kinematics to see whether they follow the stars or the gas.

Another possible scenario is that the cool gas responsible for the
emission indeed originates from the progenitors of the post-AGB
stars. They have expanded so much that they no longer appear as
planetary nebulae, but still are not as far as a randomly positioned
cloud. At this point, they are completely dispersed in the
interstellar medium and are carried away by the motion of the hot gas
and thus they appear kinematically decoupled from the stars. In this
picture, the cold gas has an internal origin, but their kinematics is
driven by the hot gas, which is kinematically decoupled from the
stars, perhaps due to mergers and the collisional nature of the gas.

Another possible solution is that the stars have a distribution that
is better resemebled by a thick disk than a sphere. Compared to the 
spherical symmetric distribution we assumed, a flatter disky distribution 
would bring the stars closer to the gas, 
raising the ionization parameter. We can investigate this by looking at whether
the stronger line emitting systems are preferentially more disky in morphology.
We leave this for future investigation.

A final possible solution is that abundance of post-AGB stars is much
larger than predicted.

To summarize, the post-AGB star photionization model can naturally
produce the general variation of ionization parameter with radius,
including the sharp rise at small radius and gentle decline on large
radius. It is also able to produce the overall direction of the
luminosity-dependence of the line ratio gradient. This result strongly
indicates that the spatial distribution of the true ionizing source is
similar to the stellar distribution. However, based on our current
knowledge about post-AGB stars, the ionization parameter they produce
would be too small, even though they may have sufficient total
luminosity. The uncertainty in the number density of post-AGB stars is
still too large \citep{Brown00, Brown08, Weston10} and observations
are too scarce. Deeper observations and larger surveys of them are
necessary to settle these questions.

\subsubsection{Other possible distributed photoionizing sources}

Low-mass X-ray binaries and extreme horizontal branch stars are two
other evolved populations that could provide some additional ionizing
photons. However, \cite{Sarzi10} have argued that they would produce
much fewer photoionizing photons than post-AGB stars. Thus they are
unlikely to be responsible on their own for the observed line
emission, or to make up the ionization parameter deficit found above.

Recently, high-mass X-ray binaries (HMXBs) and ultraluminous X-ray sources (ULXs)
(ULXs) have also been invoked to explain the LINER-like emission \citep{McKernan11}.
However, we do not think this population would solve the deficit either. 
First, in an old galaxy like those in our sample, there would be very few high-mass
X-ray binaries, because they are only associated with young stellar populations. 
Second, both HMXBs and ULXs are X-ray bright so they should have 
been included in the accounting by \cite{Eracleous10}, who showed that
extrapolating the X-ray luminosity in the nuclear region of LINERs to the ultraviolet does 
not yield enough ionizing photons to produce the nuclear \hal\ 
luminosity observed. Therefore, these components would make at most 
a minor contribution.

The hot X-ray emitting gas is also a distributed ionizing source that
can produce LINER-like emission \citep{VoitD90,DonahueV91}. Because
the hot gas density approximately follows the square root of the
stellar density, the X-ray emission should have the same luminosity
density profile as the stars. Therefore, it can also produce the
expected trend in line ratio gradient and luminosity dependence. 

However, the X-ray gas is unlikely to produce enough ionizing photons.
For the typical galaxy in our sample (the 25\% passive red galaxy at
$0.09<z<0.1$ with the brightest total line luminosity), with a median
stellar mass of $7.5\times10^{10}M_\odot$ and
$L_B=2.8\times10^{10}L_\odot$, the X-ray luminosity from the hot gas
is on the order of $10^{41} {\rm erg~s^{-1}}$ \citep{O'Sullivan01},
much lower than the total ionizing luminosity of post-AGB stars
($\sim10^{42} {\rm erg~s^{-1}}$). Therefore, they should be
subdominant to post-AGB stars and would have an even lower
contribution to the ionization parameter.

\subsection{Fast radiative shocks} 

Shocks are prevalent in many astrophysical phenomena, such as
supernova explosions, stellar winds, AGN jets and outflows, cloud
collisions, etc. Collisional excitation in the post-shock medium can
produce line ratios similar to LINERs. The fast radiative shock could
also photoionize the precursor, unshocked regions. When combined with
the emission lines produced in the cooling zone of the shock, the
resulting line ratios are similar to those of Seyferts
\citep{GrovesDS04II}.  In the section, we investigate whether the
shock model can reproduce the trends we observed in the data.

Because the shock-only model produces a better match to the LINER-like
line ratios observed in these passive red galaxies, we only consider
this model. In this model, the line ratios are determined by four
parameters: shock velocity, magnetic field strength, density, and
metallicity. Because the strong dependence of line ratios on shock
velocity, if there are a wide range of shock velocities present in a
galaxy, we should observe different widths for different lines in the
shock-only model. The data do show different widths for different
emission lines. The issue is whether the velocity dependence of line
ratios can produce the observed width difference in the right
direction.

In Section~\ref{sec:linewidth}, we showed that the \oiii\ line is on average wider than \sii\ lines by 16\%.  This means that those high-velocity line-emitting regions have a higher \oiii/\sii\ ratios than low-velocity regions. Although the bulk-motion velocity of clouds is not the same as the shock velocity, we expect in general faster moving clouds in a galaxy would generate faster shocks when they collide. If the lines are mostly produced in post-shock cooling zones resulting from cloud collisions, to explain the data would require higher \oiii/\sii\ ratio to be produced in faster shocks.   

\begin{figure}
\begin{center}
\includegraphics[totalheight=0.6\textheight,viewport=10 0 440 750, clip]{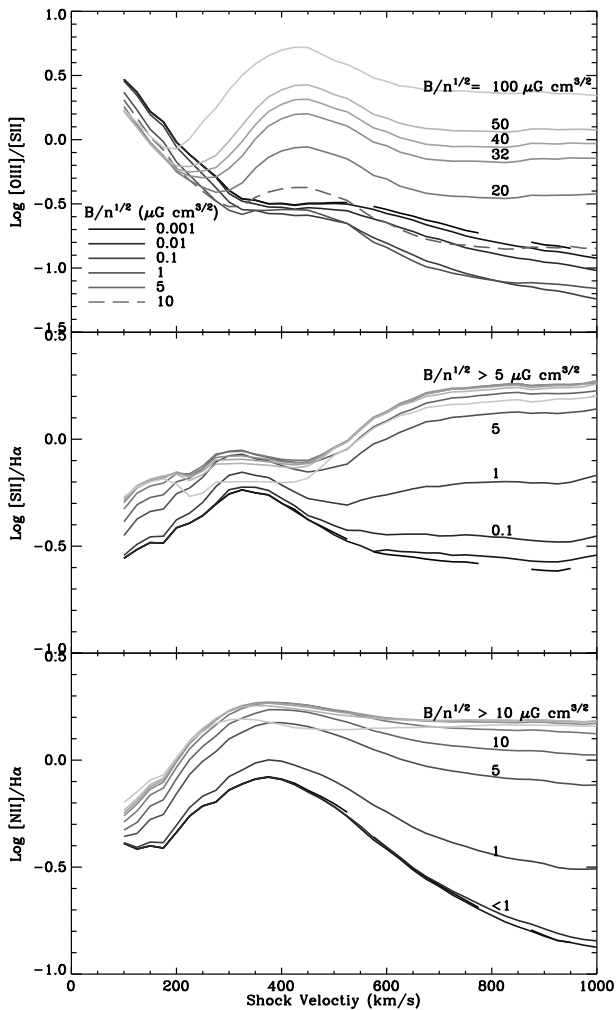}
\caption{Line ratios produced by shocks as a function of shock velocity for different magentic field strengths. The curves are plotted from dark black to light grey in order of increasing magnetic field strength.}
\label{fig:shock_velocity}
\end{center}
\end{figure}

We look at the line ratio dependence on velocity in the fast shock
models given by \cite{Allen08}, which are run using the MAPPINGS III
code.  Figure~\ref{fig:shock_velocity} shows the \oiii/\sii\ ratio as
a function of shock velocity for different magnetic field
strengths. This model is run with solar metallicity and a pre-shock
gas density of $n=100 {\rm cm}^{-3}$. In most part of the parameter
space, \oiii/\sii\ decreases with increasing velocity.  Only for
$B/n^{1/2} \ge 10 \mu{\rm G cm^{3/2}}$ and for shock velocity between
250 and 400{\rm km/s}, does the \oiii/\sii\ ratio increase with
velocity.  However, these parameter ranges do not produce the right
velocity dependence for the other line ratios. The \sii\ lines in our
sample is narrower than \hal\ lines by 7\% on average, and the
\nii\ lines are having roughly the same width as \hal\ lines. This
requires the \sii/\hal\ to decrease with velocity and \nii/\hal\ to
stay constant with velocity.  However, the model predicts a strongly
increasing \nii/\hal\ and a slightly increasing/flat \sii/\hal\ with
velocity in those particular parameter ranges, inconsistent with the
requirement to explain the observations.  Therefore, although shocks
certainly exist in most galaxies, they are probably not the dominant
source in producing the extended line emission in these passive red
galaxies.

Our conclusion agrees with the conclusion of \cite{Sarzi10}, who 
also argued against the shock scenario as the dominant
ionizing source based on the low circular velocity and low velocity 
dispersion observed, lack of morphological correlation between line 
emission structure and line ratio structure, and the flat EW distribution. 

On the other hand, \cite{Annibali10} argued that shocks could be important in
the central regions of some early-type galaxies as the AGN jet-driven outflows
or accretion onto a massive black hole could possibly reach the high velocities 
(300-500 km/s) required by the shock models. Shocks certainly exist in 
these situations, but having the right 
condition for shocks to occur does not mean shocks are directly
responsible for the ionization of the gas. Further proof, 
such as a correlation between line ratio and velocity, is necessary. 

In a totally different case, in ultraluminous infrared galaxies (ULIRGs),
shocks could indeed be responsible for producing the strong LINER-like 
emission found there \citep{Monreal-Ibero06,Monreal-Ibero10,Alonso-Herrero10}. 
As these galaxies are usually results of major mergers, stronger 
and faster shocks are more prevalent. Star formation in these galaxies 
might also be partially responsible for the line emission. 

\section{Conclusions}

In this paper, we studied the spatial distribution of LINER-like line
emission in passive red galaxies by comparing the nuclear emission
luminosity measured from the Palomar survey with the larger aperture
data from SDSS. We find strong evidence for line ratio gradients.  We
also find that different emission lines have different velocity
widths, in contrast to the uniform velocity widths in star-forming
galaxies.  We have reached the following conclusions.

\begin{enumerate}

\item In the majority of line-emitting red galaxies, the line emission
  is spatially extended and its intensity peaks at the center. The
  average \hal\ surface brightness profile can be well approximated by
  a power-law with an index of $-1.28$.  Line-emitting red galaxies
  identified with nuclear aperture spectroscopy or those with extended
  aperture spectroscopy are essentially the same population.

\item Line ratio gradients exist in these line-emitting red galaxies,
  with the very center having generally larger \nii/\hal,
  \sii/\hal, and smaller \oiii/\sii\ than the outskirts. The
  \oiii/\sii\ gradient requires an increasing ionization parameter
  towards larger distances. Because the cool gas density is likely
  to fall with radius at a much slower rate than $r^{-2}$, an outward
  increasing ionization parameter strongly disfavors AGN as the
  dominant ionizing mechanism in these galaxies.

\item The line ratio gradient can be produced by ionizing sources that
  are distributed like the stars.  This model also predicts different
  line ratio gradient trends in bright and faint galaxies, which are
  generally matched by observations. 

\item The leading candidate for the ionizing source is the population
  of post-AGB stars. The majority of these stars cannot be central
  stars of planetary nebulae, but have to be naked post-AGB stars
  creating a diffuse ionizing field.  However, the ionization
  parameter produced by post-AGB stars falls short of the required
  value by more than a factor of 10. Either the abundance of post-AGB
  stars is underpredicted or their spatial distribution has to be much
  closer to the gas clouds than assumed. The latter possibility would
  suggest a common origin of the gas and the post-AGB stars.

\item Different emission lines in passive red galaxies often have
  different widths. The \oiii\ is on average wider than \sii\ by 16\%;
  \nii\ and \hal\ are wider than \sii\ by $\sim8\%$. The width ratios
  do not vary as a function of aperture size. This latter result
  strongly suggests that the width ratio is not produced by the
  combination of the line ratio gradient and rotation, but more likely
  due to a multiphase ISM in these galaxies.

\item We considered shock models for producing these trends. Because
  line ratios produced in the cooling zone of the shocks have a strong
  dependence on the shock velocity, these models naturally produce
  width differences among different lines. However, their velocity
  dependence generate opposite width differences from what we observe.
  Therefore, shocks are strongly disfavored by our results as the
  dominant ionizing source in these passive red galaxies. However, it
  may be responsible for producing LINER-like emission found in ULIRGs.

\item The systematically different \nii/\hal\ ratio profiles between bright 
  and faint galaxies (Fig.~\ref{fig:n2ha_logu_all})
  suggest that the gas-phase metallicity is dependent on galaxy luminosity.

\end{enumerate}

Our result strongly disfavors AGN as the dominant ionization mechanism
for the line emission in passive red galaxies. However, it does not
mean that all LINERs have nothing to do with AGN. For a large fraction
of those nuclear LINERs identified in the Palomar survey, accretion
activity probably does exist, as evidenced by the detection of compact
radio core \citep{Nagar00} and X-ray point sources in their centers
\citep{Ho01}. Our result does mean that the optical line emission in
most of them are probably powered by sources unrelated with AGN
activity. The AGN is probably significantly less luminous in line
emission than previously thought and dominates on much smaller scales
than 100~{\rm pc}. This result also helps to resolve the energy budget
problem reported for most of these ``nuclear LINERs.'' The large X-ray
and radio detection fraction may be a result of more fuel supply in
these relatively ``gas-rich" early-type galaxies.

For line emission found in apertures covering much larger area, such
as in the SDSS at $z>0.02$, the line emission is nearly always
dominated by extended emission unrelated with AGN activity. Therefore,
most studies using line emission to derive AGN bolometric luminosity
for LINER-like objects using SDSS data (e.g. \citealt{KauffmannHT03,
  Kewley06, KauffmannH09,Choi09}) or higher-$z$ surveys
\citep{Bongiorno10} probably need to have their results
re-inspected. The impact is probably most significant for objects with
$L_{\rm [OIII]} \lesssim 10^6 L_\odot $, for which we have demonstrated that the
ionizing sources are outside the nucleus. The exact threshold is also
a function of the aperture size and galaxy luminosity.

Although we only focused on passive red galaxies in our investigation,
the result should also apply to LINER-like objects among younger
red-sequence galaxies. Those red LINER-like objects with smaller
$D_n(4000)$ probably also have some weak star-forming activity
contributing to their line emission, as demonstrated in
Fig.~\ref{fig:n2ha_hal_sloanpalomar}.

The extended line emission is present in more than half of
red-sequence galaxies and is much more luminous than most
low-ionization AGNs. Based on the Palomar results, only the brightest
few percent of low-ionization AGNs have a chance of detection in large
aperture spectroscopy data.

Our result favors a distribution of ionizing sources that follows the
stars, but does not confirm post-AGB stars as the ionizing sources.
Post-AGB stars are the only source that has sufficient total energy to
produce the observed emission lines. However, they fall short in the
ionization parameter. This mystery awaits future observations to
resolve.

If the post-AGB stars are confirmed as the ionization sources, the
LINERs can provide a window onto the gas dynamics of passive red
galaxies. The total flux from post-AGB stars stay fairly constant with
the age of the stellar population, except for the first Gyr after the
starburst. In this case, our results would then indicate that
differing amounts of line emission in these galaxies is mainly an
indicator of different amounts of cool/warm gas. We could therefore
use the line strength observed to study the cooling and heating of
warm gas in early-type galaxies.

This study shows the huge amount of information we can learn from wide
wavelength range, well-calibrated, high-resolution spectroscopy. It
also demonstrates the power of large statistical samples. More
detailed, spatially resolved IFU studies of nearby early-type galaxies
are obviously the next step to confirm our results. An important
lesson from this work is that the inclusion of \nii, \hal, and \sii\ in
the resolved spectra was essential to constraining the ionizing
sources; indeed, in our case the broad wavelength range is arguably
more important than the spatial resolution available to IFU
observations. This result motivates the use of IFU techniques with
broad wavelength coverage to maximize the available information.

\acknowledgements

We would like to thank the referee for detailed and thorough comments, 
which helped us improve the paper. RY would like to thank Guangtun Zhu 
and Timothy Heckman for
illuminating discussions that greatly improved this work. RY and MB
acknowledge the support of the NSF Grant AST-0908354, NASA Grant
08-ADP08-0019m, NASA Grant 08-ADP08-0072, and a Google Research Award.

Funding for the Sloan Digital Sky Survey (SDSS) has been provided by
the Alfred P. Sloan Foundation, the Participating Institutions, the
National Aeronautics and Space Administration, the National Science
Foundation, the U.S. Department of Energy, the Japanese
Monbukagakusho, the Max Planck Society, and the Higher Education Funding
Council for England. The SDSS Web site is http://www.sdss.org/.
The SDSS is managed by the Astrophysical Research Consortium (ARC) for
the Participating Institutions. The Participating Institutions are The
University of Chicago, Fermilab, the Institute for Advanced Study, the
Japan Participation Group, The Johns Hopkins University, Los Alamos
National Laboratory, the Max-Planck-Institute for Astronomy (MPIA),
the Max-Planck-Institute for Astrophysics (MPA), New Mexico State
University, the University of Pittsburgh, Princeton University, the
United States Naval Observatory, and the University of Washington.

\bibliographystyle{apj}
\bibliography{apj-jour,astro_refs}

\end{document}